\documentclass[a4paper,11pt]{article}
\pdfoutput=1 

\usepackage{jheppub} 

\usepackage[T1]{fontenc} 

\usepackage{pgfplots}
\pgfplotsset{compat=newest}
\usepgfplotslibrary{fillbetween}

\usepackage{amssymb,amsmath}
\usepackage{bbm}
\usepackage{mathtools}
\usepackage{graphicx}
\usepackage{epsfig}
\usepackage{epstopdf}
\usepackage{setspace}
\usepackage{dsfont}
\usepackage{latexsym}
\usepackage{tikz}
\usepackage{psfrag}
\usepackage{booktabs}
\usepackage{bbm}
\usepackage[toc]{appendix}
\usepackage{color}
\usepackage{datetime}
\usepackage{tensor}
\usepackage{lscape}
\usepackage{lipsum}
\usepackage{caption}
\usepackage{subcaption}

\usepackage{tikz}
\usetikzlibrary{arrows,shapes}
\usetikzlibrary{trees}
\usetikzlibrary{patterns}
\usetikzlibrary{matrix,arrows} 				
\usetikzlibrary{positioning}				
\usetikzlibrary{calc,through}				
\usetikzlibrary{decorations.pathreplacing}  
\usepackage{pgffor}							
\usetikzlibrary{decorations.pathmorphing}	
\usetikzlibrary{decorations.markings}
\tikzset{
    vector/.style={decorate, decoration={snake}, draw},
	provector/.style={decorate, decoration={snake,amplitude=2.5pt}, draw},
	antivector/.style={decorate, decoration={snake,amplitude=-2.5pt}, draw},
        smallvector/.style={decorate, decoration={snake,amplitude=1.5pt,post length=0.5mm}, draw},
    fermion/.style={draw=black, postaction={decorate},
        decoration={markings,mark=at position .55 with {\arrow[draw=black]{>}}}},
    fermionbar/.style={draw=black, postaction={decorate},
        decoration={markings,mark=at position .55 with {\arrow[draw=black]{<}}}},
    fermionnoarrow/.style={draw=black},
    gluon/.style={decorate, draw=black,
        decoration={coil,amplitude=4pt, segment length=5pt}},
    scalar/.style={dashed,draw=black, postaction={decorate},
        decoration={markings,mark=at position .55 with {\arrow[draw=black]{>}}}},
    scalarbar/.style={dashed,draw=black, postaction={decorate},
        decoration={markings,mark=at position .55 with {\arrow[draw=black]{<}}}},
    scalarnoarrow/.style={dashed,draw=black},
    electron/.style={draw=black, postaction={decorate},
        decoration={markings,mark=at position .55 with {\arrow[draw=black]{>}}}},
    bigvector/.style={decorate, decoration={snake,amplitude=4pt}, draw},
    arrow/.style={draw=black, postaction={decorate},
        decoration={markings,mark=at position 1 with {\arrow[draw=black]{>}}}},
}

\tikzstyle{block} = [draw, rectangle, 
    minimum height=3em, minimum width=6earticlem]

\definecolor{darkblue}{rgb}{0.2, 0, 0.8}

\numberwithin{equation}{section}

\usepackage{comment}

\newcommand{\reef}[1]{(\ref{#1})}

\newcommand{\be}{\begin{equation}}
\newcommand{\ee}{\end{equation}}

\def\be{\begin{equation}}
\def\ee{\end{equation}}
\def\bea{\begin{Eqarray}}
\def\eea{\end{Eqarray}}
\def\ba{\begin{array}}
\def\ea{\end{array}}
\def\bd{\begin{displaymath}}
\def\ed{\end{displaymath}}

\def\tr{{\rm tr}}

\def\a{\alpha}

\def\d{\delta}
  
\def\g{\gamma}

\def\G{\Gamma}

\def\>{\rangle} 
\def\<{\langle} 
\def\Dsl{D \hskip-.6em \raise1pt\hbox{$ / $ } }
\def\to{\rightarrow}

\newcommand{\eps}{\epsilon}

\newcommand{\gap}{\text{gap}}

\title{Flattening of the EFT-Hedron:
Supersymmetric Positivity Bounds and the Search for String Theory}

\author{Justin Berman$^a$,}
\author{Henriette Elvang$^a$,}
\author{Aidan Herderschee$^b$}

\affiliation{${}^a$ Leinweber Center for Theoretical Physics, Randall Laboratory of Physics,\\ \phantom{${}^a$} The University of Michigan, Ann Arbor, MI 48109-1040, USA
}

\affiliation{${}^b$ Institute for Advanced Study, Einstein Drive, Princeton, NJ 08540, USA}

\emailAdd{jdhb@umich.edu, elvang@umich.edu, aidanh@ias.edu}

\abstract{
We examine universal positivity constraints on $2 \to 2$ scattering in 4d planar $\mathcal{N}=4$ supersymmetric Yang-Mills theory with higher-derivative corrections. 
We present numerical evidence that the convex region of allowed Wilson coefficients (the ``EFT-hedron'') flattens completely along about one-third of its dimensions 
when an increasing number of constraints on the spectral density from crossing-symmetry are included. 
Our analysis relies on the formulation of the positivity constraints as a linear optimization problem, which we implement using two numerical solvers, SDPB and CPLEX. Motivated by the flattening, we propose a novel partially resummed low-energy expansion of the $2 \to 2$ amplitude.  As part of the analysis, 
we provide additional evidence in favor of the conjecture \cite{Huang:2020nqy} that the Veneziano amplitude is the only amplitude compatible with both S-matrix bootstrap constraints and string monodromy.
}

\begin{document} 
 \begin{flushright}
{\tt LCTP-2315} \\
\end{flushright}
\maketitle
\flushbottom


\section{Introduction}\label{s:intro}

In the framework of effective field theory (EFT),  high-energy physics can be encoded into local higher-derivative operators. If the UV theory is known, the massive degrees of freedom can be integrated out to determine the Wilson coefficients of these operators in the EFT description. 
In contrast,  from a purely low-energy perspective that ignores any details about the UV physics,
the coefficients can take on any values.
However, even without knowing specific details of the UV theory, such as its spectrum or couplings, fundamental physical principles --- locality, unitarity, and suitable assumptions about the high-energy behavior of the full scattering amplitude --- impose non-trivial constraints on the allowed values of the Wilson coefficients. 
These basic high-energy assumptions facilitate a dispersive representation of the Wilson coefficients which implies that they must lie in a convex region  sometimes called the ``EFT-hedron'' \cite{Arkani-Hamed:2020blm}. 
The allowed region of Wilson coefficients has been explored recently for theories of massless particles, including for scalars \cite{Caron-Huot:2020cmc,Arkani-Hamed:2020blm,Chiang:2021ziz},  massless pions \cite{Guerrieri_2021,Zahed:2021fkp,Albert:2022oes,Fernandez:2022kzi,Albert:2023jtd,he2023bootstrapping}, photons \cite{Alberte:2020bdz,Henriksson:2021ymi,Haring:2022sdp,deRham:2022sdl}, and gravitons \cite{Chowdhury:2021ynh,Caron-Huot:2022ugt}.

In this paper, we derive universal bounds on 4d planar $\mathcal{N}=4$ super Yang-Mills (SYM) theory with higher-derivative corrections. This is done using the  $2 \to 2$ scattering amplitude, assuming a weak coupling approximation that allows us to suppress loops of the massless SYM states. 
Using the 4-point supersymmetry Ward identities, we show that the low-energy expansion of the $2 \to 2$ color-ordered amplitude must take the form\footnote{Our  4-point Mandelstam variables are
$s= -(p_1+p_2)^2$, $t = -(p_1+p_3)^2$, and $u = -(p_1+p_4)^2$, treating all momenta as outgoing.}
\begin{equation}
\label{introansatz}
A[zz\bar{z}\bar{z}] =
-\frac{s}{u}+s^{2}
\sum_{k=0}^\infty \sum_{q=0}^k  a_{k,q} \, s^{k-q}\, u^{q} \, ,
\end{equation}
where $z$ and $\bar{z}$ are conjugate scalars of the massless $\mathcal{N}=4$ SYM spectrum and maximal supersymmetry requires $a_{k,k-q}=a_{k,q}$ (``SUSY crossing symmetry''). The $a_{k,q}$ are in 1-1 correspondence with the coefficients of the local single-trace $\mathcal{N}=4$ higher-derivative 4-field operators. 

Additionally assuming a mass-gap and a Froissart-like bound, we derive a dispersive representation for the Wilson coefficients $a_{k,q}$ for all $k,q$. We also derive two types of sum rules (or ``null constraints'') resulting from a supersymmetric version of crossing symmetry. Including both sum rules in the analysis gives optimal bounds for
any finite cutoff $k_\text{max}$ on Mandelstam terms in the low-energy ansatz \reef{introansatz} and a
spin cut-off $\ell_\text{max}$. The choice of $k_\text{max}$ corresponds to including all local $\mathcal{N}=4$ SUSY compatible 4-point operators of the schematic form $\tr(D^{2k+4}z^2\bar{z}^2)$ with $k \le k_\text{max}$ in the analysis. For example, $k_\text{max}=8$ includes operators with up to 20 derivatives in the analysis. The higher $k_\text{max}$, the stronger the bounds tend to be.

The dispersive representation of the $a_{k,q}$'s  allow us to derive bounds on ratios of Wilson coefficients. For example, we find that  $a_{0,0}$, the coefficient of the $\mathcal{N}=4$ supersymmetrization of 
$\tr F^4 \sim \tr(D^{4}z^2\bar{z}^2)$, must be bigger than any other Wilson coefficient, so it is natural to bound ratios  $\bar{a}_{k,q} \equiv a_{k,q}/a_{0,0}$. The full allowed region is then a convex subregion within the hypercube $0 \le \bar{a}_{k,q}  \le 1$. Although we do not utilize the analytic EFT-hedron bounds of \cite{Arkani-Hamed:2020blm,Chiang:2021ziz}, we still loosely refer to the allowed region as the ``$\mathcal{N}=4$ supersymmetric EFT-hedron'' or simply the ``EFT-hedron''.

To determine the bounds on (projections of) the supersymmetric EFT-hedron, we formulate the constraints as a linear optimization problem. We use two well-established linear programming solvers to numerically determine these bounds. One program is the semi-definite programming code, SDPB, developed by Simmons-Duffin for the purpose of the conformal bootstrap \cite{Simmons-Duffin:2015qma,Landry:2019qug}. The second solver is CPLEX, a commercial code maintained by IBM \cite{cplex2009v12}.  SDPB has previously been used for positivity bounds, see for example \cite{Caron-Huot:2020cmc,Chiang:2021ziz,Albert:2022oes,Albert:2023jtd}, but to our knowledge this is the first time CPLEX is used in the context of the S-matrix bootstrap. The main purpose of using both methods is to have non-trivial checks on, and comparisons of, the numerical results. We find excellent agreement between the two methods. CPLEX runs faster for the precision needed to illustrate the large-scale bounds of the allowed regions in plots, but when high-precision results are needed, SDPB is the more efficient and reliable choice. Most of the plots in the paper were generated with SDPB.

The study of the $\mathcal{N}=4$ supersymmetric EFT-hedron is partially motivated by string theory. 
Specifically, the Veneziano tree amplitude for massless Type-I open superstring scattering must be contained within the $\mathcal{N}=4$ SUSY EFT-hedron.\footnote{With the $1/\sqrt{\alpha'}$ as the mass gap, the Veneziano amplitude corresponds to a single point in the supersymmetric EFT-hedron. Unlike the string loop-amplitudes, the string tree amplitude is not sensitive to details of the compactification from 10d to 4d.}
Investigating this space may shed light on the unique properties of string theory. One long-term goal is to explore what fundamental conditions isolate the open string as the only viable UV completion of low-energy $\mathcal{N}=4$ SYM or, more generally, of even just YM theory, at tree-level. 

As a step in this direction, the authors of 
\cite{Huang:2020nqy} studied the combination of EFT-hedron bounds with the string monodromy relations \cite{Plahte:1970wy,Stieberger:2009hq,Bjerrum-Bohr:2009ulz,Bjerrum-Bohr:2010mia,Bjerrum-Bohr:2010pnr},
\begin{align}
\label{intromono}
    0=A[2134]+e^{i\pi \alpha' s}A[1234]+e^{-i\pi\alpha' t}A[1324] \,.
\end{align}
When imposed on the low-energy ansatz \reef{introansatz}, the monodromy relations \reef{intromono} fix particular linear combinations of the Wilson coefficients $a_{k,q}$, while an infinite set of coefficients remain unfixed, e.g.~at the lowest orders $a_{1,0}$, $a_{3,0}$, and  $a_{4,1}$.  The authors of \cite{Huang:2020nqy} showed that when combined with the EFT-hedron bounds of \cite{Arkani-Hamed:2020blm}, 
$a_{1,0}$ and $a_{3,0}$ were numerically fixed to be within about a percent of the string values and $a_{4,1}$ within about $50\%$. They went on to propose that string monodromy, together with positivity bounds, would isolate the open string. 

As part of our analysis, we revisit the monodromy+EFT-hedron proposal of \cite{Huang:2020nqy} and extend the results up to 20th derivative order. 
Using SDPB (along with some CPLEX cross-checks), we show that
these additional constraints now bring $a_{1,0}$ and $a_{3,0}$ to within less than $0.01\%$ of their string values. More generally, we find that the allowed regions for the coefficients unfixed by monodromy become tiny islands around the open string values. The islands continue to shrink as $k_\text{max}$ is increased. This leads to the expectation that the islands will reduce to a point in the limit of  $k_\text{max} \to \infty$. 

Working to $k_\text{max}=8$, we find the following bounds on the eight lowest Wilson coefficients left unfixed by monodromy relations:
\begin{equation}
\begin{array}{rcll}
\text{{\bf SDPB}}
&\!\!\text{{\bf bounds}}
&~~&
\text{{\bf String Value}} ~a_{k,q}^{\text{str}}
\\
1.201982&\leq a_{1,0}\leq & 1.202061 \quad & a_{1,0}^{\text{str}}=\zeta_{3} \approx 1.202057 \\
1.036923&\leq a_{3,0}\leq& 1.036937 \quad 
& a_{3,0}^{\text{str}}=\zeta_{5} \approx 1.036928 \\
0.04053 &\leq a_{4,1}\leq& 0.04063 \quad & 
a_{4,1}^{\text{str}}=\tfrac{3}{4} \zeta_6 -\tfrac{1}{2}
\zeta_3^2 \approx 0.04054 \\
1.0083481 &\leq a_{5,0}\leq &1.0083495 \quad & 
a_{5,0}^{\text{str}}=\zeta_{7} \approx 1.0083493 \\
0.008649 &\leq 
a_{6,1}\leq &0.008729 \quad &
a_{6,1}^{\text{str}} = \tfrac{5}{4} \zeta_8 -
\zeta_3 \zeta_5 \approx 0.008651 \\
1.00200830 &\leq 
a_{7,0}\leq &1.00200891 \quad & a_{7,0}^{\text{str}}=\zeta_{9} \approx 
 1.00200839 \\
0.00031 &\leq 
a_{7,2}\leq &0.00041 \quad & 
a_{7,2}^{\text{str}}=-\tfrac{7}{4} 
\zeta_6 \zeta_3 
+ \tfrac{1}{6} \zeta_3^3 - \tfrac{9}{4} \zeta_4 \zeta_5 
- 3 \zeta_2 \zeta_7 
+ \tfrac{28}{3} \zeta_9 \approx 0.00032 \\
0.00203 &\leq a_{8,1}\leq &0.00212 \quad & a_{8,1}^{\text{str}}=
\tfrac{7}{4}  \zeta_{10} - \tfrac{1}{2} \zeta_5^2 - \zeta_3 \zeta_7 \approx 0.00204\,. \\
\end{array}
\end{equation}
Going to higher $k_\text{max}$ to get even stronger bounds is in principle straightforward and just requires more computation time. We find that the bounds shrink toward zero as a power-law (or faster) in $k_\text{max}$, so this supports the proposal of \cite{Huang:2020nqy} that string monodromy combined with positivity bounds single out the Veneziano amplitude.

The string monodromies impose linear relations among the Wilson coefficients. From a geometric perspective,  in the space of $a_{k,q}$'s these relations define a higher-dimensional ``plane''; we call this space the {\em monodromy plane}. Meanwhile, at finite $k_\text{max}$, the positivity bounds give an allowed region, the supersymmetric EFT-hedron, which has  co-dimension zero in the space of SUSY crossing-symmetric Wilson coefficients. When monodromy and positivity isolate a small island at finite $k_\text{max}$, that is the statement that the monodromy plane and the supersymmetric EFT-hedron intersect each other in a small volume. The claim that this small volume of allowed values of Wilson coefficients shrinks to a point with increasing $k_\text{max}$ is then the statement that the monodromy plane intersects the
supersymmetric EFT-hedron at a single point in the limit $k_\text{max} \to \infty$.
That one point has the values of the Wilson coefficients corresponding to the Veneziano amplitude.

\begin{figure}[t!]
\begin{center}
\begin{tikzpicture}
\centering
\node (image) at (0,0){\includegraphics[width=0.8\textwidth]{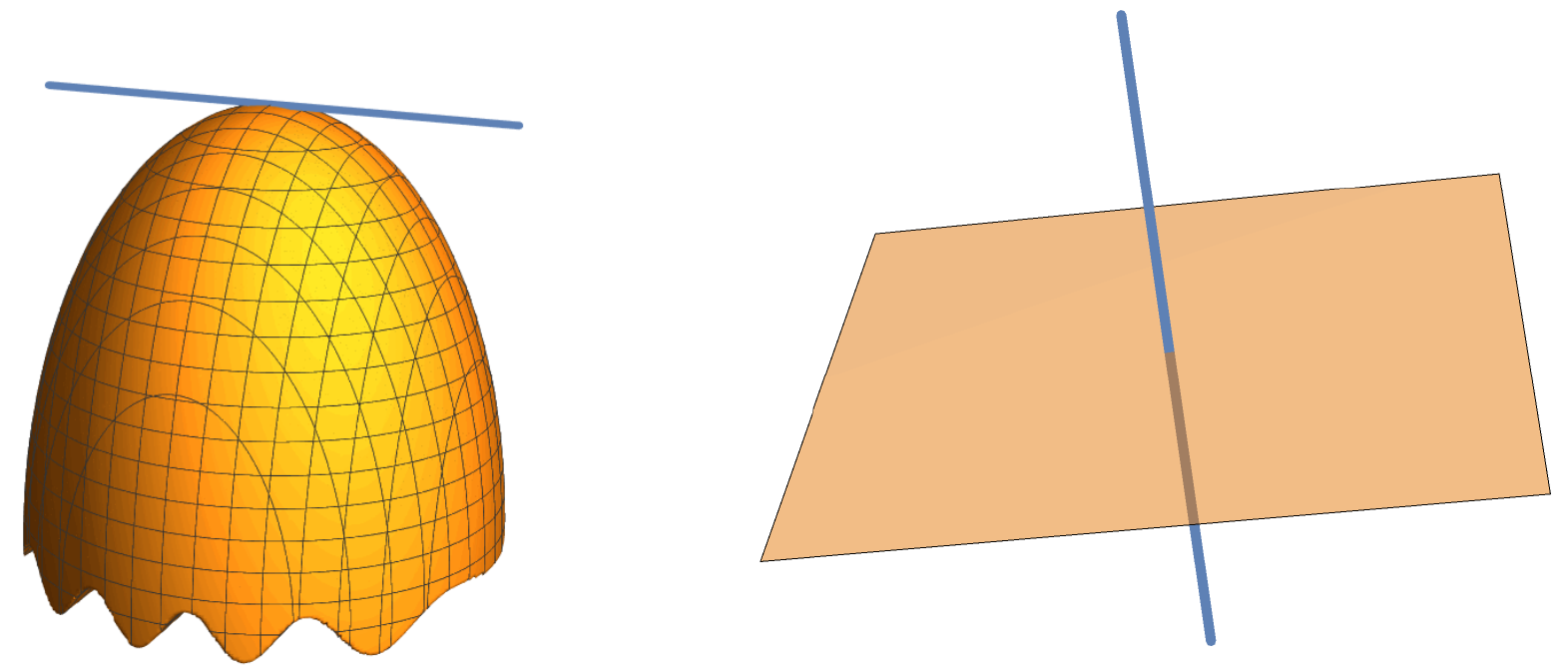}};
\fill[red] (-3.9,1.78) circle (2pt);
\fill[red] (2.975,-0.15) circle (2pt);
\end{tikzpicture}
\end{center}
\caption[cartoon]{3D cartoon of how the monodromy line (blue) could intersect with an EFT-hedron of codimension zero (left) or nonzero (right).}
\label{fig:cartoon}
\end{figure}

The intersection at a point could happen in two distinct ways, as illustrated at the cartoon level in Figure \ref{fig:cartoon}: either the monodromy plane is tangent to the allowed EFT-hedron region in the limit $k_\text{max} \to \infty$ or the monodromy plane intersects the interior of the finite-$k_\text{max}$ EFT-hedron in a manner such that, as $k_\text{max} \to \infty$, the EFT-hedron flattens, leading to a point of intersection between the two spaces. 

One of the core results of this paper is numerical evidence for the conjecture that the allowed space of Wilson coefficients flattens out to a space of lower dimensionality in the limit of 
$k_\text{max} \to \infty$; i.e.~that the second of the two geometric options in Figure \ref{fig:cartoon} is correct. 
This flattening conjecture implies that in the $k_\text{max} \to \infty$ limit, all theories are boundary theories and there are much stronger constraints among Wilson coefficients than one might naively have anticipated. 

To show this, we move the monodromy plane around so that it intersects the EFT-hedron at different points determined by randomly generated linear combinations of known allowed models. If
the first picture in Figure \ref{fig:cartoon} is correct, doing so would lead to islands that do not continue to shrink with increasing $k_\text{max}$. What we find is that these allowed islands do indeed continue to shrink. 

\begin{figure}[t]
\begin{center}
\begin{tikzpicture}
	\node (image) at (0,0) {\includegraphics[width=0.5\textwidth]{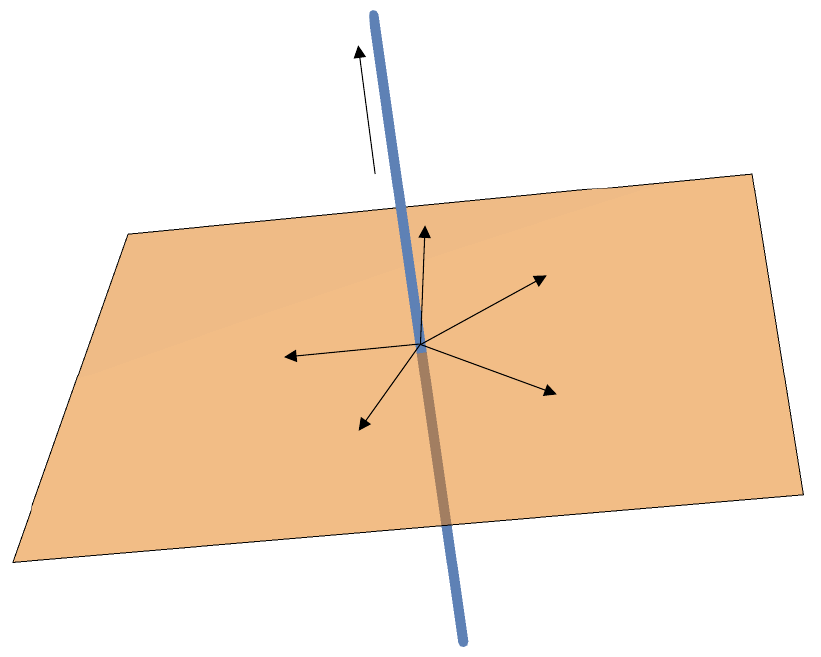}};
        \node at (-0.9,2.1) {$\bar{A}_i^{(k)}$};
        \node at (0.6,0.8) {$\bar{r}_{0}^{(0)}$};
        \node at (1.7,0.6) {$\bar{r}_{1}^{(2)}$};
        \node at (1.8,-0.6) {$\bar{r}_{2}^{(2)}$};
        \node at (-0.6,-1.3) {$\bar{r}_{3}^{(3)}$};
        \node at (-1.35,-0.15) {$\bar{r}_{4}^{(4)}$};
        \node at (-1.0,0.3) {.};
        \node at (-0.9,0.4) {.};
        \node at (-0.8,0.5) {.};
        \fill[red] (0.08,-0.15) circle (2pt);
\end{tikzpicture}
\end{center}
\caption{A simplified sketch  the parameterization of the flattened EFT-hedron parameterized by monovariables 
$\bar{r}_{i}^{(k)} = {r}_{i}^{(k)}/a_{0,0}$
 and its transverse directions 
 $\bar{A}_{i}^{(k)} = {A}_{i}^{(k)}/a_{0,0}$.}
\label{fig:newparam}
\end{figure}
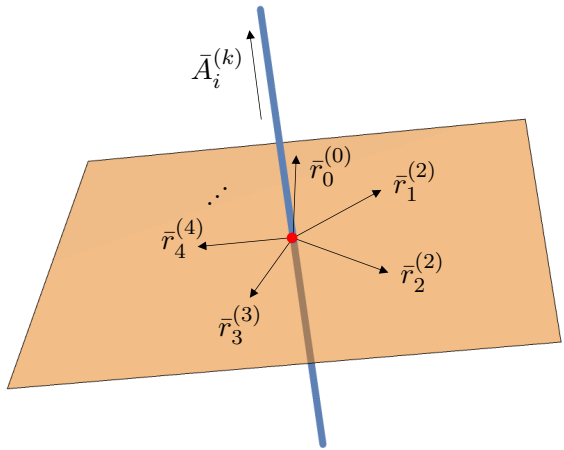

Specifically, we find evidence that fixing two-thirds of the Wilson coefficients and imposing locality, unitary, and the Froissart bound is sufficient to fix the remaining one-third of Wilson coefficients. 
Thus, the flattening suggests that
there is a ``better'' low-energy representation of the EFT amplitude than the standard one in \reef{introansatz}, perhaps even one in which certain combinations of Mandelstam polynomials have been resummed. The parameters should split up into two sets: those corresponding to coordinates along the flattened EFT-hedron (we call these {\em monovariables} $r^{(k)}_i$) and those transverse to it, $A_{i}^{(k)} = a_{1,0}, a_{3,0}, a_{4,1}, a_{5,0},a_{6,1}$, etc. This is illustrated in Figure \ref{fig:newparam}. 
Our analysis suggests that we rewrite the EFT amplitude as
\be \label{newexpA4}
  A[zz\bar{z}\bar{z}] = - \frac{s}{u}
  + s^2
  \bigg( 
   \sum_{k,i}  r_i^{(k)} P^{(k)}_i(s,u)
   +
  \sum_{k,i} A_{i}^{(k)} Q^{(k)}_i(s,u)
   \bigg) \,,
\ee
where $P^{(k)}_i(s,u)=P^{(k)}_i(u,s)$ are specific symmetric degree-$k$ polynomials in $s$ and $u$. 
The $Q^{(k)}_i(s,u)=Q^{(k)}_i(u,s)$ are infinite sums of $s$,$u$ symmetric polynomial terms whose lowest-order terms are degree $k$. 
The key point of flattening is the claim that for any choice of monovariables $r^{(k)}_i$ in the EFT-hedron, the positivity constraints of the S-matrix bootstrap uniquely fix all coefficients $A^{(k)}_i$. 
At large $k$, the monovariables $r^{(k)}_i$ account for two-thirds of all the variables: thus one only needs to specify two-thirds of all the EFT coefficients to know the whole low-energy expansion. Importantly, we also find evidence that there does exist a form of the amplitude \reef{newexpA4} in which the $Q^{(k)}_i(s,u)$ can be resummed. The answer is surprisingly simple and of the form
\be
\label{resummed}
\sum_{k,i} A_{i}^{(k)} Q^{(k)}_i(s,u) 
=
\frac{\sin(\pi t)}{\pi}\sum_{k,i} \tilde{A}_{i}^{(k)} \mathcal{S}^{(k)}_i(s,t,u)\,,
\ee 
where $t = -s-u$ and $\mathcal{S}^{(k)}_i$ represents the degree-$k$ Mandelstam polynomials that are fully symmetric in $s,t,u$. The coefficients $\tilde{A}_{i}^{(k)}$ are finite linear combinations of the $A_{i}^{(k)}$. 
We have verified this ansatz up to 20th order in the Mandelstam variables.\footnote{Some readers may recognize the RHS of \reef{resummed} as an ansatz that trivially solves the string monodromy relations without restricting the coefficients $\tilde{A}_{i}^{(k)}$. We discuss this in Section \ref{s:goodcoords}.}

Let us come back to the statement that monodromy and positivity constraints combine to single out the open string tree amplitude. Since the string monodromy relations arise from the worldsheet description of the string, it seems dissatisfactory to impose them in order to isolate the Veneziano amplitude. However, monodromy relations can be shown \cite{Chen:2022shl} to arise also in purely field theoretic context, namely from scalar bi-adjoint (BAS) effective field theory. The BAS EFT appears in the tree-level double-copy where it can be used as a way to generate the higher-derivative corrections to other theories; relevant for us here is the double-copy relation
\be\label{N4dcEFT}
 \text{($\mathcal{N}=4$ SYM EFT)} =
 \text{(BAS EFT)}
\otimes_\text{FT}
 \text{(pure $\mathcal{N}=4$ SYM)} \,,
\ee
The conjecture of \cite{Chen:2022shl} is that the most general tree amplitude (S)YM EFT obtained by the double-copy \reef{N4dcEFT} automatically satisfies the string monodromy relations.  
The results obtained here, expanding on the earlier results of \cite{Huang:2020nqy}, then leads to the conjecture that {\em among all the 4-point $\mathcal{N}=4$ SYM EFT tree amplitudes obtained from the double-copy \reef{N4dcEFT}, the unique one compatible with unitarity, locality, the existence of a mass-gap, and the Froissart bound is the Veneziano open-string tree amplitude.}

The above conjecture brings the assumption of monodromy constraints a step down toward a more purely low-energy effective field theory approach. It would of course be very interesting to have assumptions that are even more fundamental than the existence of the EFT double-copy relation \reef{N4dcEFT} and that is a goal of future work. 

Finally, let us note that with SDPB, it is also possible to extract the spectrum of theories that lie at the boundaries of the allowed space.\footnote{We thank Jan Albert and David Poland for useful discussions related to this topic.} One might expect that when we numerically fix Wilson coefficients to be close to their string values, the spectrum of the extremal theory closely mirrors that of string theory. We find that the spectra for these extremal theories do match some of the leading Regge trajectories, but that there are also spurious states that do not match the open string spectrum. These presumably disappear at higher $k_\text{max}$ and $\ell_\text{max}$. We leave a more detailed analysis of these numerical spectra to the future.

{\bf The paper is organized as follows.} We start in Section \ref{s:susy} by deriving the constraints of $\mathcal{N}=4$ supersymmetry on the $2 \to 2$ scattering amplitude. Next, in Section \ref{s:disp}, we state the technical assumptions, then derive the dispersive representation of the Wilson coefficients as well as null constraints on the spectral density from SUSY crossing symmetry.
In Section \ref{s:opti} we formulate the optimization problems and briefly discuss the implementations in SDPB and CPLEX. Readers familiar with the dispersive arguments may choose to skip ahead to the core results. 

In Section \ref{s:allowedR}, we explore some of the simplest bounds and offer brief comparisons of SDPB and CPLEX. The main takeaway from this section is there is no sign that the Veneziano amplitude should lie at a kink or any other particular feature of the bounds in these projections.

The analysis with monodromy imposed as additional null constraints is presented in Section \ref{sted}. In Section \ref{s:flat}, we change the monodromy constraints to ``monovariable constraints'' and present numerical evidence supporting the conjecture that the supersymmetric EFT-hedron flattens in the $k_\text{max} \to \infty$ limit.
We also introduce the novel partially-resummed parameterization of the low-energy amplitude.  
We conclude with a discussion and future outlook in Section \ref{s:discuss}. The Appendix contains technical discussions of the numerical implementation. 

\vspace{2mm}
\noindent {\bf Note added:} while preparing this paper, we became  aware of partially overlapping results  in \cite{yutinetal} in which the authors   find even stronger bounds on the $a_{1,0}$, $a_{3,0}$, and $a_{4,1}$ Wilson coefficients when the string monodromy relation is imposed.

\section{Amplitudes in $\mathcal{N}=4$ SYM + h.d.} 
\label{s:susy}

In this section, we derive an ansatz for the low-energy expansion of the $2\rightarrow 2$ scattering amplitudes in $\mathcal{N}=4$ supersymmetric Yang-Mills EFT with gauge group $SU(N)$ in the strict large-$N$ limit. We provide some examples of different UV completions that give non-zero Wilson coefficients in the low-energy theory. 

\subsection{$\mathcal{N}=4$ Superamplitude} 
\label{s:superA}

The massless  
$\mathcal{N}=4$ vector supermultiplet consists of 16 states: two gluon helicity states $g^\pm$, four pairs of positive and negative helicity gluinos $\lambda^{A}$ and $\lambda^{ABC}$, and three pairs of complex scalars $z^{AB}$. The on-shell states transform in fully antisymmetric irreducible representations of the $SU(4)_R$ R-symmetry group;  $A,B,C = 1,2,3,4$ are R-indices. 

The scattering amplitudes of an 
$\mathcal{N}=4$ SYM EFT can be encoded into on-shell superamplitudes. At 4-point, we write
\be
  \label{superA4}
  \mathcal{A}_4 = 
  \delta^{8}(\tilde{Q}) 
  \frac{[12]^2}{\<34\>^2} \,f(s,u) 
  ~~~~\text{with}~~~~
  \delta^{8}(\tilde{Q}) 
  = \frac{1}{2^4} \prod_{A=1}^4
  \sum_{i,j = 1}^4 
  \<ij\> \eta_{iA} \eta_{jA} \,.
\ee
The ordering of the external states is understood to be 1234 unless otherwise specified. The on-shell superspace formalism with the Grassmann variables $\eta_{iA}$ can be found in 
 Chapter 4 of \cite{Elvang:2013cua,Elvang:2015rqa}. To project out component amplitudes from the superamplitude, one takes  derivatives with respect to the Grassmann variables $\eta_{iA}$ to match the R-indices of the $i$th state. A positive helicity gluon corresponds to the $SU(4)_R$ singlet with no indices, whereas the negative helicity gluon corresponds to the singlet with all four R-indices, i.e.~$g^- = g^{1234}$. Thus, projecting out the 4-gluon amplitude from \reef{superA4} gives
\be  
  \label{A4gluons}
  A[++--] = [12]^2 \<34\>^2 f(s,u)\,,
\ee
where $\pm$ is shorthand for the gluon helicity states. 
In pure (S)YM theory, the tree-level Parke-Taylor gluon amplitude is
\be
 A^\text{YM}[++--]
 = \frac{\<34\>^4}{\<12\>\<23\>\<34\>\<41\>}
= - \frac{[12]^2 \<34\>^2}{su}\,,
\ee
so $f(s,u) = -1/(su)$ in pure (S)YM. 

Consider a pair of conjugate scalars $z = z^{12}$ and $\bar{z} = z^{34}$ of the massless $\mathcal{N}=4$ supermultiplet. 
Projecting out three different 4-scalar amplitudes from the superamplitude \reef{superA4}, we find
\be
  \label{twoA4s}
  A[z z\bar{z} \bar{z} ]
  = s^2 f(s,u)\,,
  ~~~
  A[z\bar{z} z\bar{z} ]
   = t^2  f(s,u)= 
A[\bar{z} z\bar{z} z]
   \,.
\ee
Cyclicity requires $A_4[2341]=A_4[1234]$, so, together with the supersymmetry requirement $A_4[z\bar{z} z\bar{z}]=A_4[\bar{z} z\bar{z} z]$ from \reef{twoA4s}, we see that $f$ must be symmetric in $s$ and $u$:
\be
  \label{fus}
  f(u,s) = f(s,u)\,.
\ee
We call this equality ``crossing symmetry''. It is clearly satisfied by the Parke-Taylor amplitude, but it must hold for the full amplitude as well.   

\subsection{Low-Energy Ansatz} 
\label{loweanz}
On-shell, local, higher-derivative operators are in 1-1 correspondence with polynomial terms in $f(s,u)$,  subject to momentum conservation $s+t+u = 0$. Hence, in the low-energy expansion, the most general form\footnote{We exclude pole terms $(u/s + s/u)$ or $(1/s + 1/u)$ in $f(s,u)$ because by \reef{twoA4s} they would imply that $A[zz\bar{z}\bar{z}]$ has residues of order $s^2$ and $s^3$ 
in the $u$-channel  corresponding to exchanges of massless spin 2 and 3 states. Alternatively, one can argue the absence of these pole terms by the fact that there exist no $\mathcal{N} = 4$ SUSY compatible 3-point interactions made from the $\mathcal{N} = 4$ SYM fields.}
of the 4-point amplitude is
\be
  \label{fsu}
  ~~f(s,u) = - \frac{1}{su} + \sum_{0 \le q\leq k} a_{k,q}s^{k-q}u^{q}\,.
\ee
This assumes a weak-coupling limit in which we exclude contributions from loops of massless particles which would generate logarithms in the low-energy ansatz \reef{fsu} and  running of the EFT couplings.

Of particular interest to us is the 4-scalar amplitude $A[zz\bar{z}\bar{z}]$. By \reef{twoA4s}, the most general ansatz for this component is
\begin{equation}
\label{ansatz}
\boxed{\phantom{\Bigg[}
A[zz\bar{z}\bar{z}] =
-\frac{s}{u}+s^{2}\sum_{0 \le q\le  k} a_{k,q} \, s^{k-q}\, u^{q} \, .~~}
\end{equation}
Since not all higher-derivative operators are compatible with $\mathcal{N}=4$ supersymmetry, the 
coefficients $a_{k,q}$ are restricted. Specifically, the crossing relation \reef{fus} requires us to impose
\be
\label{crossingsymmet}
\boxed{\phantom{\bigg[}
  \text{Crossing / SUSY:}~~~~
  a_{k,k-q} = a_{k,q}\,
  ~~~\text{for all}~~ 0 \le q \le k \,.}
\ee
The $a_{k,q}$ are Wilson coefficients for (linear combinations of) the on-shell local operators compatible with supersymmetry.
The factor of $s^2$ multiplying the sum in \reef{ansatz} means that no interaction with less than four derivatives contributes to this amplitude, i.e.~there are no $\mathcal{N}=4$ compatible interactions of the form $\tr(z^2\bar{z}^2)$ and $\tr(D^2 z^2\bar{z}^2)$.\footnote{$\mathcal{N}=4$ SYM does, of course, have local 4-scalar interactions, but these have a different R-symmetry index structure, for example $z^{12}z^{23}z^{34}z^{41}$; i.e.~they involve two different pairs of conjugate scalars, not just one.} This is simply the statement that $\tr(F^4)$ is the lowest-dimensional $\mathcal{N}=4$ supersymmetric higher-derivative operator available in the vector sector. Indeed, 
$a_{0,0}$ is the coefficient of $\tr(F^4)$, 
$a_{1,0}=a_{1,1}$ is the coefficient of the unique $\mathcal{N}=4$ SUSY compatible operator $\tr(D^2F^4)$, etc.

\subsection{Examples}
\label{s:ampsEx}
Here, we present relevant examples of amplitudes compatible with the SUSY crossing constraint \reef{fus}.

\subsubsection{Veneziano Amplitude} \label{sec:vamp}
The Veneziano amplitude for tree-level scattering of massless open superstrings is unitary \cite{Maity:2021obe,Arkani-Hamed:2022gsa} and compatible with  $\mathcal{N}=4$ supersymmetry upon restriction to 4d. Projecting to two pairs of massless external scalars, the Veneziano amplitude is 
\begin{align}
\label{veneziano}
A^{\text{str}}[zz\bar{z}\bar{z}] = -(\a's)^2\frac{\G(-\a's)\G(-\a'u)}{\G(1-\a'(s+u))}.
\end{align}
Expanding in small $\alpha's$ and $\alpha'u$ we find
\be
\label{VenezianoExp}
A^{\text{str}}[zz\bar{z}\bar{z}]
= -\frac{s}{u}
+ s^2 \bigg(
 \zeta_2 \alpha'^2 
+ \zeta_3 \alpha'^3 (s+u)
+ \zeta_4 \alpha'^4 
\big(s^2 +  u^2\big)
+ \frac{1}{4}\zeta_4 \alpha'^4
s u 
+ \ldots
\bigg)\ ,
\ee
where $\zeta_s$ is the Riemann zeta function. We can read off
\begin{align}
\label{akq-veneziano}
a_{0,0}^{\text{str}} = \zeta_2 \,\alpha'^{2}\,,
~~~
a_{1,0}^{\text{str}} = a_{1,1}^{\text{str}} = \zeta_3 \,\alpha'^{3}\,,
~~~ a_{2,0}^{\text{str}} = a_{2,2}^{\text{str}}
=\zeta_4  \,\alpha'^{4}\,,
~~~
a_{2,1}^{\text{str}} = \tfrac{1}{4}\zeta_4 \, \alpha'^{4} \,,
~~\text{etc.}
\end{align}
from the comparison to the general ansatz \reef{ansatz}. 

\subsubsection{1-loop Contribution from the Coulomb Branch}
\label{s:coulomb}

Consider one-loop contributions from BPS states on the Coulomb branch as an example of a UV completion \cite{Craig:2011ws,Kiermaier:2011cr,Herderschee:2019dmc}.\footnote{We thank Enrico Hermann for suggesting this example.} We start with  $\mathcal{N}=4$ SYM  with a $SU(N')$ gauge group and go onto the Coulomb branch such that the gauge symmetry is broken to $SU(N)\times SU(N'')$ with $N'=N+N''$. We restrict the external states to be massless states transforming in the adjoint of the $SU(N)$ sector. 
The massive states that transform in the (anti-)fundamental of $SU(N)$ and (anti-)fundamental of $SU(N'')$ couple quadratically to the massless external states and therefore start contributing only at 1-loop order.

The loop contributions of the massive states of $\mathcal{N}=4$ SYM on the Coulomb branch do not include bubble or triangle integrals 
(see for example
\cite{Boels:2010mj} and \cite{Abhishek:2023lva}), so the only contribution is from box-diagrams. The explicit contribution of a single massive BPS state with mass $m$ running in the loop is
\begin{equation}
A^{\textrm{1-loop}}[zz\bar{z}\bar{z}]=\frac{6s^{2}}{\pi^{2}}\int \frac{d^{4}\ell}{[s_{\ell}-m^{2}][s_{\ell,1}-m^{2}][s_{\ell,12}-m^{2}][s_{\ell,123}-m^{2}]} \,.
\end{equation}
This box-diagram was shown in  \cite{Davydychev:1993ut} to be given by an Appell’s hypergeometric function of two variables, $F_3$:
\begin{equation}
\begin{split}
A^{\textrm{1-loop}}[zz\bar{z}\bar{z}]&=\frac{s^2}{m^4} F_3\Big( 1,1,1,1;\frac{5}{2} \Big| \frac{s}{4m^2},\frac{u}{4m^2}\Big), \\[1mm]
&=\frac{s^2\Gamma(5/2)}{m^4} \sum_{j,l=0}^{\infty}\frac{\Gamma(1+l)\Gamma(1+j)}{\Gamma(5/2+j+l)}\left ( \frac{s}{4m^{2}} \right )^{j}\left ( \frac{u}{4m^{2}} \right )^{l} \,.
\end{split}
\end{equation}
The Wilson coefficients are  
\begin{align}
 &
 a_{0,0}
 = \frac{1}{m^4}\,,~~~
 a_{1,0}
 =a_{1,1}
 = \frac{1}{10}\frac{1}{m^6}\,,~~~
 a_{2,0}
 =a_{2,2}
 = \frac{1}{70}\frac{1}{m^8}\,,
 ~~~
 a_{2,1}
 = \frac{1}{140}\frac{1}{m^8}\,,
 \nonumber \\
 &
 a_{3,0}
 =a_{3,3}
 = \frac{1}{420}\frac{1}{m^{10}}\,,
 ~~~
 a_{3,1}
 =a_{3,2}
 = \frac{1}{1260}\frac{1}{m^{10}}\,,
 ~~~\text{etc.}
\end{align}
Note that we have dropped overall factors in the box-diagram and tuned the normalization of the amplitude to make $a_{0,0}=1/m^4$. The dispersive representation bounds ratios of Wilson coefficients, so the overall scaling does not matter.

\subsubsection{Infinite Spin Tower}
\label{s:otherAmps}
Another amplitude that satisfies the crossing constraint \reef{crossingsymmet} is 
\be
A^{\text{IST}}[zz\bar{z}\bar{z}] = -\frac{s}{u}+\frac{s^2}{\left(m^2- s\right)\left(m^2- u\right)}\,.
\label{higherspinmass}
\ee
The coefficients of the low-energy expansion are 
\be
 a_{k,q} = \frac{1}{m^{2k+4}}
 ~~~
 \text{for all}
 ~
 k,q\,.
 \label{bthy2}
\ee
The $A^{\text{IST}}$-amplitude tends to show up as an allowed solution in S-matrix bootstrap analyses \cite{Caron-Huot:2020cmc,Albert:2022oes}. However, it has an unsuppressed infinite tower of higher spin states, all with the same mass, so it is not expected to arise from a physical theory even though it is not explicitly forbidden by our assumptions. 

\section{Dispersive Representation}
\label{s:disp}

We study the full color-ordered $\mathcal{N}=4$ SYM EFT scalar amplitude 
\be
  \label{Asu}
  A(s,u)=A[zz\bar{z}\bar{z}] \,,
\ee
with supersymmetry constraints and a low-energy expansion as discussed in the previous section. In this section, we exploit the expected analytic structure of the amplitude to derive positivity bounds for the Wilson coefficients $a_{k,q}$ of the lower-energy expansion. We summarize the technical assumptions in Section \ref{s:assumptions}, then derive 
the dispersive representation of the Wilson coefficients $a_{k,q}$ in Section \ref{sec:derivbds}. The final result is given in equation (\ref{finalequ}), and the most basic consequences are discussed in Section \ref{s:basic}. In Section \ref{s:nullconsstraints} we derive additional ``null constraints'' on the  Wilson coefficients.

\subsection{Assumptions} \label{s:assumptions}

 We make the following set of assumptions:
\begin{enumerate}
    \item The gauge group has large rank, so we can work in the large-$N$ limit. This ensures that the color-ordered amplitude \reef{Asu} has no $t$-channel poles or discontinuities.
    \item The theory admits a weak coupling description. This means that we can ignore loops of massless particles and take the low-energy expansion of the amplitude to be \reef{ansatz}.
    \item The theory has a mass gap, $M_\text{gap}$, such that there are no states with nonzero mass below $M_\text{gap}$.
    \item The amplitude admits a  partial wave decomposition \begin{equation}\label{partialwavdec}
    A(s,u)=16\pi \sum_{\ell=0}^{\infty}(2\ell+1)\, a_{\ell}(s)\,P_{\ell}\big(\cos(\theta)\big)\,,    
    \end{equation}
    where $\cos(\theta)=1+2u/s$ and the Legendre polynomials $P_{\ell}$ are labeled by the spin $\ell$. Crucially, unitarity requires  $\textrm{Im}\big(a_{\ell}(s)\big)\ge 0$.\footnote{$\textrm{Im}(a_{\ell}(s))$ is also bounded from above, but we do not impose the upper bound in our analysis.}
    \item For fixed $u<0$ and sufficiently large $|s|$, the amplitude is analytic in $s$ away from the real axis in the complex $s$-plane.
    \item The amplitude obeys a Froissart-Martin-like bound: 
    \begin{equation}
    \begin{split}
    \label{matrinlikeboundss}
    &\textrm{fixed $u<0$:}~~ \lim_{s\rightarrow \infty} \frac{A(s,u)}{s^{2}}=0\,,   \\[1mm]
   &\textrm{fixed $t<0$:}~~ \lim_{s\rightarrow \infty} \frac{A(s,-s-t)}{s^{2}}=0    \ .
   \end{split}
    \end{equation}
\end{enumerate}

A rigorous derivation of Property 5 for general theories is not currently known, but it does hold at all orders in perturbation theory \cite{Caron-Huot:2020cmc,Arkani-Hamed:2020blm,Camanho:2014apa}. Property 6 can be shown to hold with assumptions about the UV behavior of the theory \cite{Heisenberg:1949kqa,Froissart:1961ux}: it was argued in \cite{Arkani-Hamed:2020blm} that if the amplitude is analytic and polynomially bounded as $A(s,u)<s^{N}$ for any $N$ at large $s$, then (\ref{matrinlikeboundss}) follows from unitarity. 

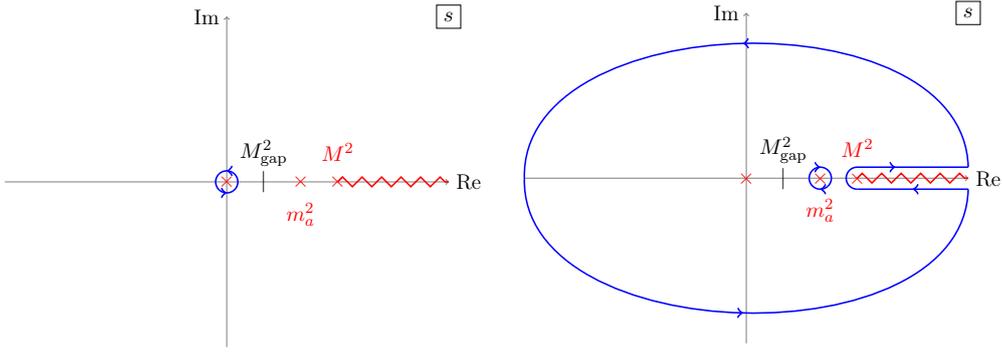
\begin{figure}[t]
	\centering
    \begin{subfigure}[t]{0.45\textwidth}
     \centering
	\raisebox{2mm}{\scalebox{0.73}{
  \begin{tikzpicture}
  [
    decoration={%
      markings,
      mark=at position 0.5 with {\arrow[line width=1pt]{>}},
    }
  ]
  \draw [help lines,->] (-4,0) -- (4,0) coordinate (xaxis);
  \draw [help lines,->] (0,-3) -- (0,3) coordinate (yaxis);
  \node [right] at (xaxis) {Re};
  \node [left] at (yaxis) {Im};
  
  \path [draw, line width=0.8pt, postaction=decorate,blue] (0.2,0) arc (0:180:.2);
  \path [draw, line width=0.8pt, postaction=decorate,blue] (-0.2,0) arc (-180:0:.2);

  \node at (2/3,0) {$|$};
  \node[above=0.15cm] at (2/3,0) {$M_{\gap}^{2}$};

  \node[red] at (1+1/3,0) {$\times$};
  \node[red] at (0,0) {$\times$};
  \node[red] at (2,0) {$\times$};

  \draw[line width=0.8pt, red] (2,0) [decorate, decoration=zigzag] --(4,0);

  \node[below=0.25cm,red] at (1+1/3,0) {$m_{a}^{2}$};
  \node[above=0.25cm,red] at (2,0) {$M^2$};

 \node[draw] at (4,3) {$s$};

\end{tikzpicture}
}}
    \end{subfigure}\hspace{0mm}\begin{subfigure}[t]{0.45\textwidth}
\centering
\scalebox{0.73}{
\begin{tikzpicture}
  [
    decoration={%
      markings,
      mark=at position 0.5 with {\arrow[line width=1pt]{>}},
    }
  ]
  \draw [help lines,->] (-4,0) -- (4,0) coordinate (xaxis);
  \draw [help lines,->] (0,-3) -- (0,3) coordinate (yaxis);
  \node [right] at (xaxis) {Re};
  \node [left] at (yaxis) {Im};

  \node[red] at (1+1/3,0) {$\times$};
  \node[red] at (0,0) {$\times$};
  \node[red] at (2,0) {$\times$};

  \node at (2/3,0) {$|$};
  \node[above=0.15cm] at (2/3,0) {$M_{\gap}^{2}$};

  \draw[line width=0.8pt, red] (2,0) [decorate, decoration=zigzag] --(4,0);
  
  \path [draw, line width=0.8pt, postaction=decorate,blue] (0.8+1/3,0) arc (180:0:.2);
  \path [draw, line width=0.8pt, postaction=decorate,blue] (1.2+1/3,0) arc (0:-180:.2);
  
  \node[below=0.25cm,red] at (1+1/3,0) {$m_{a}^{2}$};
  \node[above=0.25cm,red] at (2,0) {$M^2$};
  
  \path [draw, line width=0.8pt, postaction=decorate,blue,bend angle=90] (4,0.2) to[bend right] (-4,0);
 \path [draw, line width=0.8pt, postaction=decorate,blue,bend angle=90] (-4,-0) to[bend right] (4,-0.2);
  
  \path [draw, line width=0.8pt, postaction=decorate,blue] (4,-0.2) -- (2,-0.2);
  \path [draw, line width=0.8pt, postaction=decorate,blue] (2,-0.2) arc (270:90:0.2) -- (4,0.2 );
  
  \node[draw] at (4,3) {$s$};
\end{tikzpicture}
}
   \end{subfigure} 
   \caption{\label{fig:contour_deformation}
The contour deformation that converts  (\ref{formaldefinition}) to (\ref{discontdefor}). The contribution from the arc at infinity vanishes due to Property 6. The contour around the branch cut can be identified with the discontinuity of the $s$-channel branch-cut. We only include a single simple pole explicitly in the figure, but there can be an infinite number of massive simple poles on the real positive $s$-axis.}
\end{figure}

\subsection{Dispersive Representation of Wilson Coefficients} \label{sec:derivbds}

Each of the Wilson coefficients in the low-energy expansion \reef{ansatz} of $A(s,u)$ can be extracted by the contour integral 
\begin{equation}\label{formaldefinition}
a_{k,q}=\frac{1}{q!}\frac{\partial^q}{\partial u^{q}}\int_{\mathcal{C}^{\star}} \frac{ds'}{2\pi i} \frac{A(s',u)}{s'^{k-q+3}} \bigg|_{u=0}  \ , 
\end{equation}
where the contour $\mathcal{C}^{\star}$ is a small circle surrounding  $s=0$ in the complex $s$-plane. 
The  ``+3'' in the power of $s'$ in the denominator accounts for the factor of $s^2$ in the low-energy ansatz \reef{ansatz}. Together with the assumption (\ref{matrinlikeboundss}), the ``+3'' ensures that the contour deformation described in Figure~\ref{fig:contour_deformation} has vanishing contribution from the contour at infinity 
for any $0 \le q \le k$. Therefore, we get
\begin{equation}\label{discontdefor}
a_{k,q}=\frac{1}{q!}\frac{\partial^q}{\partial u^{q}} \left (\frac{1}{\pi}\int \frac{ds'}{s'^{k-q+3}}\,\textrm{Im}A(s',u)\right )\Bigg|_{u=0} \,
\end{equation}
for all $0 \le q \le k$. 
Here we used that the discontinuity of the amplitude is proportional to its imaginary part,\footnote{For simplicity of the presentation, we have absorbed single-particle contributions into the definition of the discontinuity. The single particle contributions can be treated separately, as done in Refs.~\cite{Arkani-Hamed:2020blm,Caron-Huot:2020cmc}, but they are eventually absorbed into the spectral function and make no practical difference for the final form of the dispersive representation.}  $2i$Im$[A(s,u)] = A(s+i\eps,u)-A(s-i\eps,u)$. There are no $t$-channel contributions because we work in the planar limit and no $u$-channel contributions because we work at fixed $u<0$. 

Next, we use the partial wave decomposition, 
\begin{equation}\label{imagpartam}
\textrm{Im}(A)
=
16\pi\sum_{\ell=0} (2\ell+1)\,\textrm{Im}(a_{\ell}(s'))\,P_{\ell}\bigg(1+\frac{2u}{s'}\bigg) \ .    
\end{equation}
The Legendre polynomials can be written 
\begin{equation}
P_{\ell}\big(1+2\delta\big)=\sum_{q=0}^{\ell}v_{\ell,q}
\delta^{q} 
\quad 
\text{with}~~
v_{\ell,q}=\frac{\prod_{a=1}^{q} \big[\ell(\ell+1)-a(a-1)\big]}{(q!)^{2}}\,,
\end{equation}   
 where $v_{\ell,q} \ge 0$ for $\ell \ge q$ and we define $v_{\ell,q} = 0$ for $q > \ell$.
Since the only dependence on $u$ enters \reef{discontdefor} via the Legendre polynomials, 
taking $q$ $u$-derivatives and then setting $u=0$ picks out the coefficient $v_{\ell,q}$. Hence, after a change of integration variable, $s' = M^2$, \reef{discontdefor} becomes 
\begin{equation}\label{projectivepro}
a_{k,q}=\sum_{\ell=0}\int_{M^{2}_{\textrm{gap}}}^\infty dM^{2}\, \rho_{\ell}(M^{2})\left  ( \frac{1}{M^{2}}\right )^{k+3}v_{\ell,q}  \ ,
\end{equation}
where 
\begin{equation}\label{definrho}
\rho_{\ell}(M^{2})=16  (2\ell+1)\,\textrm{Im}\big(a_{\ell}(M^{2})\big)    \ .
\end{equation}
Unitarity requires $\rho_{\ell}(M^{2}) \ge 0$ and this places non-trivial restrictions on the $a_{k,q}$. 

It is useful to rewrite \reef{definrho} in terms of dimensionless quantities.\footnote{Equivalently, we could set $M_{\gap} = 1$.} 
To make the Wilson coefficients dimensionless, we multiply \reef{projectivepro} by $(M_{\gap}^{2})^{(k+2)}$ and redefine the  $a_{k,q}$ as
\begin{equation}\label{WCredef}
(M_{\gap}^{2})^{(k+2)} a_{k,q}\to a_{k,q}\ .
\end{equation}
We then define 
\be\label{redefex}
x=\frac{M_{\textrm{gap}}^{2}}{M^{2}} ~~~\text{and}~~~  
p_{\ell}(x)= x\,\rho_\ell\big(M_{\textrm{gap}}^{2}/x\big)\geq 0\, 
\ee
in terms of which \reef{projectivepro} becomes
\begin{equation}
\label{finalequ}
\boxed{
~~~a_{k,q}=\sum_{\ell=0}\int_{0}^{1} 
dx\, p_{\ell}(x) \,
x^{k}\, v_{\ell,q}, ~~
\quad p_{\ell}(x) \ge 0 \ .~~~}
\end{equation}
This is the dispersive representation of the Wilson coefficients that we use to derive bounds in the following sections. Physically,   
equation (\ref{finalequ}) relates the individual low-energy Wilson coefficient to the integral over the high-energy spectrum.

\subsection{Basic Consequences} \label{s:basic}
It is immediately clear from \reef{redefex} and \reef{finalequ} that all Wilson coefficients have to be non-negative,
\be
  \label{justpos}
  a_{k,q} \ge 0 \,.
\ee
Further, since $0 \le x \le 1$ in \reef{finalequ}, we must have
\begin{equation}
\label{sbd1q}
    a_{k',q} \leq a_{k,q}\,
    ~~~
    \text{for}
    ~~~ k \leq k' \ .
\end{equation}

We can now use the crossing conditions, $a_{k,k-q} = a_{k,q}$, along with \reef{sbd1q} to see that
\begin{equation}\label{a10max}
\begin{split}
a_{0,0} \geq &~a_{1,0} \geq a_{2,0}\geq a_{3,0} \ldots\\
&~~\shortparallel\\
&~a_{1,1} \geq a_{2,1} \geq a_{3,1}\ldots\\
&~~~~~~~~~~~~~~~~~~~~\shortparallel\\
&~~~~~~~~~~~~~~~~~~~a_{3,2}\ldots\\
&~~~~~~~~~~~~~~~~~~~~~\vdots
\end{split}
\end{equation}
Thus, $a_{0,0}$ is the largest Wilson coefficient, so if $a_{0,0} = 0$, all other $a_{k,q}$'s must vanish. In other words, unless the supersymmetrization of the operator $\tr F^4$ is included, there can be no other higher-derivative operators. 

Given a set of Wilson coefficients $a_{k,q}$ with a valid a dispersive representation (\ref{finalequ}), a new set of Wilson coefficients defined by
\begin{equation}\label{projecteq}
\forall k,q:\ a_{k,q}'=\lambda a_{k,q}, \quad \lambda>0    
\end{equation}
also trivially admits a valid dispersive representation. Therefore, the bounds only apply to ratios of Wilson coefficients.
Since $a_{0,0}$ is the largest Wilson coefficient, it is natural to study the bounds on the ratios $a_{k,q}/ a_{0,0}$. Each such ratio must obey
\begin{align} \label{singleakqbd2}
0 \leq \frac{a_{k,q}}{a_{0,0}} \leq 1 \,.
\end{align}
The more detailed shape of the higher-dimensional bounded space of allowed Wilson coefficients is studied using numerical methods in the following sections. For optimal bounds, we need to incorporate additional constraints, as discussed next.

\subsection{Null Constraints}
\label{s:nullconsstraints}

When the dispersive representation \reef{finalequ} is plugged into 
the SUSY crossing condition $a_{k,q}-a_{k,k-q} = 0$, we find the following ``null constraint'' on $p_{\ell}(x)$:
\begin{equation}\label{crossingsymmimpy}
\forall ~ k,q\!:
~~~
\sum_{\ell=0}\int_{0}^{1}dx\, p_{\ell}(x)\,\mathcal{X}^{\ell,x}_{k,q} =0
~~~\text{with}~~~
\mathcal{X}^{\ell,x}_{k,q}=
x^{k}
\big[v_{\ell,q}-v_{\ell,k-q}\big]  \,.
\end{equation}
An additional null constraint arises from a version of the dispersive argument implemented for fixed $t$ rather than fixed $u$. It takes the form 
\begin{equation}
\label{crossingsymmimpy2}
\begin{split}
&\forall ~ k,q\!:
~~~ \sum_{\ell}\int_{0}^{1} dx\, p_{\ell}(x)\,\mathcal{Y}^{\ell,x}_{k,q} =0
\\
&
\text{with}~~~
\mathcal{Y}^{\ell,x}_{k,q}= x^{k}\left [v_{\ell,q}-(-1)^{\ell}\sum_{q'=0}^{k}(-1)^{q'}v_{\ell,q'}\left (\binom{q'}{k-q}+\binom{q'}{q}\right )\right ].
\end{split}
\end{equation}
We derive this relation below. It can be thought of as the supersymmetric version of the crossing symmetry sum rule found in \cite{Albert:2022oes} for the four-pion amplitude.\footnote{The sum rules in Eq. (\ref{crossingsymmimpy2}) are particular linear combinations of those given in Ref. \cite{Albert:2022oes}. For example, ${\mathcal{Y}^{\ell,x}_{2,1,l}}^{\text{ours}}=2{\mathcal{Y}^{\ell,x}_{2,0,l}}^{\text{theirs}}-{\mathcal{Y}^{\ell,x}_{2,1,l}}^{\text{theirs}}$.
}
Note that the $\mathcal{X}^{\ell,x}_{k,q}$ and $\mathcal{Y}^{\ell,x}_{k,q}$ null constraints are not all linearly independent. 
For example, at a given $k$, only the null constraints from $\mathcal{Y}^{\ell,x}_{k,q}$ with $q\leq \left \lfloor  k/2\right \rfloor$ are linearly independent when the $\mathcal{X}^{\ell,x}_{k,q}$ null constraints are imposed for all $q\leq k$. Physically, we can interpret (\ref{crossingsymmimpy}) and (\ref{crossingsymmimpy2}) as non-trivial constraints from maximal supersymmetry on the spectrum of intermediate states.

\vspace{2mm}
\noindent {\bf Derivation of the Null Constraint \reef{crossingsymmimpy2}.}\\[1mm]
The  core idea necessary to derive the null constraint \reef{crossingsymmimpy2} is that there is a fundamentally new representation of the $a_{k,q}$ when working at fixed $t$ instead of fixed $u$. 
To start, we define 
\begin{equation}
b_{k,q} =\frac{1}{q!}\frac{\partial^q}{\partial t^{q}}\int_{\mathcal{C}^{\star}} \frac{ds'}{2\pi i} \frac{A(s',-s'-t)}{s'^{k-q+3}} \bigg|_{t=0} \ .
\end{equation}
The low-energy expansion of the amplitude identifies the $b_{k,q}$ as the Wilson coefficients in the representation,
\begin{equation}
A(s,-s-t)=  \frac{s}{s+t}
+ 
s^2 \sum_{0 \le q\le  k} b_{k,q}s^{k-q}t^{q} \,,
\end{equation}
and hence the $b_{k,q}$ are related to the $a_{k,q}$ of \reef{ansatz} as
\begin{equation}\label{b2a}
a_{k,q}=\sum_{q''=q}^{k}(-1)^{q''}\binom{q''}{q}\,b_{k,q''} \ .
\end{equation}
Performing the same contour deformation as before, we find a contribution from both the $u$- and $s$-channel branch-cuts:
\be
\begin{split}
  \oint_{\mathcal{C}^{\star}} \frac{ds'}{2\pi i} \frac{A(s',-s'-t)}{s'^{k-q+3}} 
  &= \frac{1}{\pi} \int_{M_\text{gap}^2}^\infty
  ds' \, \frac{\text{Im}\, A(s',-s'-t)}{s'^{k-q+3}} 
  \\
  &- \frac{1}{\pi} 
  \int_{-\infty}
  ^{-M_u^2-t}
 ds' \,  
\frac{\text{Im}\, A(s',-s'-t)}{s'^{k-q+3}}\,,
\end{split}
\ee
where $M_\text{gap}^2$ is the start of the cut / lowest mass in the $s$-channel and $M_u^2$ is the start of the $u$-channel cut / lowest mass in the $u$-channel. We make no assumptions regarding $M_u^2$ at this stage in the calculation. For the $u$-channel cut, we use that 
\be
 A(s',-s'-t)
  = s'^2 f(s',-s'-t)
  = s'^2 f(-s'-t,s')
  = \frac{s'^2}{(s'+t)^2} 
  A(-s'-t, s')
\ee
where $\mathcal{N}=4$  supersymmetry required the crossing symmetry for $f$ in \reef{fus}. A
 change variables $s'\rightarrow -s'-t$ gives 
\be
\begin{split}
  \oint_{\mathcal{C}^{\star}} \frac{ds'}{2\pi i} \frac{A(s',-s'-t)}{s'^{k-q+3}} 
  &= \frac{1}{\pi} \int_{M_\text{gap}^2}^\infty
  ds' \, \frac{\text{Im}\, A(s',-s'-t)}{s'^{k-q+3}} 
  \\
  &- \frac{1}{\pi} 
  \int^{\infty}
  _{M_u^2}
 ds' \,  
\frac{1}{s'^{2}}\frac{\text{Im}\, A(s',-s'-t)}{(-s'-t)^{k-q+1}}\,.
\end{split}
\ee
Now the integrand in the second line is over positive $s'$ and we know that the discontinuity in the $s$-channel cannot begin below $M_\text{gap}^2$, so we can replace $M_u^2$ with $M_\text{gap}^2$.

Next, we use the partial wave expansion for $\text{Im}\, A(s',-s'-t)$ at fixed $t$,
\begin{equation}\label{partialwavdec2}
\begin{split}
    &
    A(s,-s-t)=16\pi \sum_{\ell=0}^{\infty}(-1)^{\ell}(2\ell+1)\, a_{\ell}(s)\,P_{\ell}\Big(1+\frac{2t}{s}\Big)\,,
\end{split}
\end{equation}
where we have used that $P_{\ell}(-x)=(-1)^\ell P_{\ell}(x)$. 
The dispersive representation for $b_{k,q}$ then becomes 
\begin{equation}
\begin{split}
b_{k,q}&=\frac{1}{q!}\frac{\partial^{q}}{\partial t^{q}}\bigg (\sum_{\ell=0}^{\infty}\int_{M_\text{gap}^2}^{\infty} dM^{2}\frac{(-1)^{\ell}\rho_{\ell}(M^{2})}{M^{2(k-q+3)}}P_{\ell}\left(1+\frac{2t}{M^{2}}\right)\\
&\hspace{2cm}-\sum_{\ell=0}^{\infty}\int_{M_\text{gap}^2}^{\infty} dM^{2}\frac{(-1)^{\ell}\rho_{\ell}(M^{2})}{M^{4}(-M^{2}-t)^{k-q+1}}P_{\ell}\left(1+\frac{2t}{M^{2}}\right) \bigg ) \bigg|_{t=0} \ .
\end{split}
\end{equation}
We make the  $b_{k,q}$ dimensionless by rescaling them with powers of $M_\text{gap}^2$ as in (\ref{WCredef}), and we change integration variable from $M^2$ to $x$ as in (\ref{redefex}). The  result is  independent of $M_\text{gap}$ and can be written
\begin{equation}\label{almostfinalform}
\begin{split}
b_{k,q}&=\frac{1}{q!}\frac{\partial^{q}}{\partial t^{q}}
\bigg [\sum_{\ell=0}^{\infty}\int_{0}^{1} dx\ (-1)^{\ell}\,
p_{\ell}(x)\,x^{k-q}M_{\gap}^{2q}
\Bigg ( 1-\frac{(-1)^{k-q+1}}
{\big(1+\frac{xt}{M_{\gap}^2}\big)^{k-q+1}}\Bigg ) 
P_{\ell}\Big(1+\tfrac{2xt}{M_{\gap}^2}\Big) \bigg ] \bigg|_{t=0}, \\
&=\sum_{\ell=0}^{\infty}\int_{0}^{1}dx\ p_{\ell}(x)(-1)^{\ell}x^{k}\left [v_{\ell,q}+(-1)^{k}\sum_{q'=0}^{q}(-1)^{-q'}\binom{k-q'}{q-q'}v_{\ell,q'} \right ] \,.
\end{split}
\end{equation}
Finally, we plug the dispersive representation \reef{finalequ} for $a_{k,q}$ and 
\reef{almostfinalform} for $b_{k,q}$ into \reef{b2a}. Using the binomial product identity
\begin{equation}
\sum_{q''=q}^{k}(-1)^{q''}\binom{q''}{q}\binom{k-q'}{q''-q'}=(-1)^{k}\binom{q'}{k-q}\ ,
\end{equation}
we arrive at the $\mathcal{Y}^{\ell,x}_{k,q}$ null constraints \reef{crossingsymmimpy2}. 

%

\section{Bounds as an Optimization Problem}
\label{s:opti}

The dispersive representation \reef{finalequ}, along with the null constraints, bound the region of allowed Wilson coefficients. The space is projective since we place bounds only on the ratio of Wilson coefficients. Moreover, the allowed region is convex since any positive sum of allowed coefficients much again be allowed. We  refer to the convex space of allowed coefficients as the ``supersymmetric EFT-hedron'', even though the way we determine the bounds is different from the moment map approaches in \cite{Arkani-Hamed:2020blm} and \cite{Chiang:2021ziz}. 

Since the space of Wilson coefficients has a large dimension, we typically study projections of the 
supersymmetric EFT-hedron into a plane in order to visualize the bounds. Determining optimal bounds of such projections can be formulated as an optimization problem suitable for linear and semi-definite programming as we show in Section \ref{nearoptimalbound}. We use the semi-definite program SDPB \cite{Simmons-Duffin:2015qma,Landry:2019qug} and the IBM program CPLEX \cite{cplex2009v12} to numerically compute near-optimal bounds.

\subsection{Formulation as an Optimization Problem}
\label{nearoptimalbound}

For a projection of the supersymmetric EFT-hedron to the $(a_{k,q}/a_{0,0},  a_{k',q'}/a_{0,0})$-plane, we  determine the allowed range of  $a_{k,q}/a_{0,0}$ for a given fixed value of  $a_{k',q'}/a_{0,0} = R$ subject to the null constraints  (\ref{crossingsymmimpy}) and (\ref{crossingsymmimpy2}). This is implemented by writing the dispersive representation and null constraints in a vector equation 
\begin{align}
\vec{V}&=\sum_{\ell=0}^
{\ell_{\text{max}}}\int_{0}^{1} dx \,p_{\ell}(x) \,\vec{E}_{\ell,x} \label{vectorequation} 
\end{align}
where
\begin{align}
\vec{V}=\begin{pmatrix}
a_{0,0}\\ 
a_{k,q}\\ 
a_{k',q'} - Ra_{0,0}\\
\sum_{\ell}\int dx\, p_{\ell}(x)\mathcal{Y}^{\ell,x}_{0,0} \\
\vdots\\
\sum_{\ell}\int_{0}^{1}dx\, p_{\ell}(x)\mathcal{X}^{\ell,x}_{1,0} \\
\vdots
\end{pmatrix}&, \quad ~~\vec{E}_{\ell,x}= \begin{pmatrix}
1 \\ 
x^{k} v_{\ell,q}\\ 
x^{k'} v_{\ell,q'} - R\\
1-2(-1)^{\ell}\\ 
\vdots \\
x(v_{\ell,0}-v_{\ell,1})\\ 
\vdots\\
\end{pmatrix}
\label{vectorequation2} 
\end{align}
The first two rows encode the dispersive representations of $a_{0,0}$ and $a_{k,q}$. The third row enforces the condition $a_{k',q'} = Ra_{0,0}$ as a null constraint together with all the  SUSY crossing null constraints in the fourth row and down. We include the linearly independent $\mathcal{X}^{\ell,x}_{k,q}$ and $\mathcal{Y}^{\ell,x}_{k,q}$ null constraints for all $0\le q \le k$ up to some maximum value for $k$, $k_\text{max}$; this corresponds to considering constraints from local operators in the higher-derivative expansion up to and including $2k_\text{max}+4$ derivatives. For practical implementation, the sum over spins is truncated at some maximum value $\ell_\text{max}$. The bounds we derive consequently depend on the choice of $k_\text{max}$ and $\ell_\text{max}$.

Consider the the relation  $\sum\int dx\, p_{\ell}(x) = a_{0,0}$; since all $p_{\ell}(x)$ are positive, each  $p_{\ell}(x)$ is bounded from above by $a_{0,0}$ and can only reach that value if all the other $p_{\ell}(x)$'s vanish. The geometric interpretation of \reef{vectorequation} is then that (projectively mod $a_{0,0}$) the vector $\vec{V}$ must lie inside the convex region whose vertices are determined by the $\vec{E}_i$'s. Our goal is to find the maximum allowed value of the 2nd component of $\vec{V}$ subject to the constraint of $a_{k',q'}/a_{0,0} = R$  and the null constraints.

The maximization problem can be brought to the standard form for linear optimization as follows. Introduce a vector $\vec\alpha$ of the same length as $\vec{V}$, 
\be
  \label{thealpha}
  \vec\alpha = 
   (A, \,-1, \,\alpha_3,\alpha_4, \ldots)
\ee
and dot it into \reef{vectorequation} to get
\begin{align}
\label{alphadottedin}
\vec{\a}\cdot \vec{V} =\sum_{\ell}\int dx \ p_{\ell}(x)\,\,\vec{\a}\cdot\vec{E}_{\ell,x} \,.
\end{align}
Imposing the null constraints gives  $\vec\alpha \cdot \vec{V} = A\,a_{0,0} -  a_{k,q}$. Hence, {\em if} the righthand side of \reef{alphadottedin} is positive, we get 
\begin{align}
\frac{a_{k,q}}{a_{0,0}} \leq A\ .
\end{align}
Thus, $A$ is the upper bound on allowed values of $a_{k,q}/a_{0,0}$ on the support of the null constraints.   
One can then argue that  the problem of  maximizing $a_{k,q}/a_{0,0}$ subject to the null constraints is equivalent to {\em minimizing} $A$ subject to the positivity constraints 
\begin{align}
\vec{\alpha}\cdot \vec{E}_{\ell,x}&\geq 0 ~~ \text{for all $\ell = 0,1,\ldots, \ell_\text{max}$ and $0\leq x\leq1$.}
\label{Eq:posconslprog}
\end{align}
The parameterization of $\vec\alpha$ in \reef{thealpha} is such that  the optimization of $A$ under the 
inequalities \reef{Eq:posconslprog} imposes the null constraints of \reef{vectorequation2}. 

To summarize, the linear optimization problem is: find $\vec{\alpha}$ such that $A=\vec{\alpha} \cdot (1,0,0,\ldots)$ is minimized subject to $\vec{\alpha}\cdot \vec{E}_{\ell,x}\geq 0$ for all $\ell$ up to $\ell_\text{max}$ and all $0\leq x\leq1$. The relevant part of the output $\vec{\alpha}$ is the first component $A$, because that tells us the maximally allowed value of $a_{k,q}/a_{0,0}$ subject to the null constraints. The setup \reef{vectorequation}-\reef{vectorequation2} can be adjusted to compute both upper and lower bounds on the Wilson coefficients $a_{k,q}/a_{0,0}$. Additional null constraints, such as monodromy conditions and variants thereof, can also be included; see Section \ref{sted}.

\subsection{Implementation in SDPB}\label{sec:sdpb}

SDPB takes as input a finite set of vertex vectors, $\vec{E}_{a,x'}$, labeled by the discrete index $a$. Each element of the vector is a polynomial in a variable $x'$ that is assumed to take values between zero and infinity.  SDPB numerically solves for the optimal solution $\vec{\alpha}$ subject to the positivity constraints $\vec{\alpha}\cdot \vec{E}_{a,x'}\geq 0$ for all $a$ and $x'$.

Our optimization problem is not quite of this form because our $x$ ranges from $0\leq x \leq 1$, so we define $x$ in terms if $x'$ as
\begin{equation}\label{repsdpb1}
x \equiv \frac{1}{1+x'} \,.
\end{equation}
Furthermore, because the  elements of the SDPB vertex vectors must be polynomial in $x'$, we rescale our vertex vectors as  
\begin{equation}\label{repsdpb2}
\vec{E}_{\ell,x} \rightarrow (1+x')^{k_\textrm{max}} \vec{E}_{\ell,x'} \ .
\end{equation}
This can also be thought of as simply rescaling $p_{\ell}(x)$.

Now our optimization problem can be directly implemented in SDPB. For example, suppose we are maximizing $a_{2,1}/a_{0,0}$ while fixing $a_{2,0}/a_{0,0}=R$. The corresponding $\vec{V}$ is given by 
\begin{equation}
\vec{V}=\begin{pmatrix}
a_{0,0} \\[0.0ex]
a_{2,1} \\[0.0ex]
a_{2,0}-Ra_{0,0} \\[0.5ex]
\sum_{\ell}\int_{0}^{1}dx\, p_{\ell}(x)\mathcal{Y}^{\ell,x}_{0,0} \\[0.5ex]
\sum_{\ell}\int_{0}^{1}dx\, p_{\ell}(x)\mathcal{Y}^{\ell,x}_{1,0} \\[0.5ex]
\vdots\\[0.5ex]
\sum_{\ell}\int_{0}^{1}dx\, p_{\ell}(x)\mathcal{X}^{\ell,x}_{1,0} \\[0.5ex]
\vdots
\end{pmatrix}
\end{equation}
and, specifically for $k_\text{max}=2$,  the $\vec{E}_{\ell,x}$-vectors become
\begin{equation}
\vec{E}_{\ell,x}=\begin{pmatrix}
1\\ 
x^{2}\\[0.5ex]
x^{2}v_{\ell,1}-R  \\[0.5ex]
1-2(-1)^{\ell} \\[0.5ex]
\left(1-(-1)^\ell (1-2 \ell (\ell+1))\right) x \\[0.5ex]
\vdots\\[0.5ex]
x(v_{\ell,0}-v_{\ell,1})\\[0.5ex]
\vdots
\end{pmatrix}\rightarrow \begin{pmatrix}
(1+x')^{2}\\[0.5ex]
1\\[0.5ex]
v_{\ell,1}-R(1+x')^{2}  \\[0.5ex]
(1+x')^{2}(1-2(-1)^{\ell}) \\[0.5ex]
(1+x')\left(1-(-1)^\ell (1-2 \ell (\ell+1))\right)  \\[0.5ex]
\vdots\\[0.5ex]
(1+x')(v_{\ell,0}-v_{\ell,1})\\[0.5ex]
\vdots
\end{pmatrix}
\end{equation}

In Appendix \ref{app:linprogacc}, we discuss the algorithm's sensitivity to the choice of $\ell_\text{max}$.  

\subsection{Implementation in CPLEX}\label{sec:CPLEX}

In addition to using SDPB, we also compute bounds using the linear programming solver CPLEX. 
Unlike semi-definite programming, for which we can input vectors $\vec{E}_{\ell,x}$ with a continuous variable $x$, CPLEX needs input vectors with discrete values of $x$. Therefore, we discretize the mass-spectrum in the integral over $x = M_\text{gap}^2/M^2$ in \reef{vectorequation} by selecting a set of $x_\text{max}$ values $0<x_1 < x_2 < \ldots < x_{x_\text{max}}  = 1$ and approximating the integral as a sum. We introduce a collective index $i = (x_{n_i}, \ell_i)$ that allows us to combine the sums over the mass-spectrum and the spins $\ell$, so that \reef{vectorequation} becomes 
\begin{align}\label{dissum}
    \vec{V}=\sum_{\ell=0}^
{\ell_{\text{max}}}\int_{0}^{1} dx \,p_{\ell}(x) \,\vec{E}_{\ell,x}
    ~~\to~~
    \vec{V}=\sum_{i}p_{i}\vec{E}_{i} \ .
\end{align}
Because of the mass discretization, CPLEX underestimates the bounds compared to SDPB for given $k_\text{max}$ and $\ell_\text{max}$. 
The finer the discretization (i.e.~greater values of  $x_{\max}$), the closer the CPLEX bounds are  to the SDPB bounds. We provide some representative examples in Section \ref{s:CPLEXvsSDPB}.

\section{Allowed Regions}\label{s:allowedR}
In this section, we give examples of allowed regions and compare SDPB with CPLEX. We study how the bounds depend on the number of higher-derivative operators included in the analysis. Recall that $k$ labels local $\mathcal{N}=4$ SUSY operators of the schematic form $\tr D^{2k} F^4 \sim\tr(D^{2k+4}z^2\bar{z}^2)$, so including operators with $k \le k_\text{max}$ corresponds to including scalar field operators with up to and including $2 k_\text{max}+4$ derivatives. The  $a_{k,q}$ are the Wilson coefficients, with $q$ labeling the different independent $\mathcal{N}=4$ SUSY operators at order $k$. 
For each $k_\text{max}$, the choice of upper bound on spins, $\ell_\text{max}$, is made to ensure the bounds converge as a function of $\ell_\text{max}$ to the desired numerical precision. Examples of such benchmarking are given in Appendix \ref{app:linprogacc}.

To compare with known amplitudes, such as the open string and other examples in Section \ref{s:ampsEx}, we perform the rescaling \reef{WCredef}, $a_{k,q} \to a_{k,q} M_\text{gap}^{2k+4}$, to make the 
Wilson coefficients dimensionless in units of the mass gap. 

Section \ref{s:exbounds} presents examples of bounds on the lowest-dimension Wilson coefficients, and in Section \ref{s:CPLEXvsSDPB} we compare results of SDPB and CPLEX.

\subsection{Examples}
\label{s:exbounds}

We found in Section \ref{s:basic} that $a_{0,0}$ is the largest Wilson coefficient, and it is therefore 
 natural to focus on bounds on the ratios $a_{k,q}/a_{0,0}$. To simplify the notation, we  define
\begin{align}
\bar{a}_{k,q} \equiv \frac{a_{k,q}}{a_{0,0}}
    ~~~\text{with}~~~~
    0 \le \bar{a}_{k,q} \le 1 \,.
\end{align}
To visualize the bounds on the multi-dimensional space of Wilson coefficients $\bar{a}_{k,q}$, we project onto 2-dimensional regions $(\bar{a}_{k,q},\bar{a}_{k',q'})$. In these 2d plots, the Veneziano amplitude \reef{veneziano}-\reef{akq-veneziano} with $M_\text{gap}^2 =1/\alpha'$ is shown as a {\bf red dot}. With $a_{0,0} = \zeta_2$ for the open string, the lowest $\bar{a}_{k,q}$ values are
\be
\label{abarVen}
\text{Veneziano:}~~~
\bar{a}_{1,0}=\frac{\zeta_{3}}{\zeta_2}\approx 0.73\,,~~~
\bar{a}_{2,0} = \frac{\zeta_{4}}{\zeta_2} \approx 0.66 \,,
~~~
\bar{a}_{3,0} = \frac{\zeta_{5}}{\zeta_2}\approx 0.63\,, ~~~\text{etc.}
\ee
Varying $M_\text{gap}^2 \alpha'$ between 0 and 1 gives a set 
of Wilson coefficients that must also lie in the allowed region. These values for the open string are shown as the {\bf red dashed curves} in the plots.

The Coulomb branch 1-loop amplitude from Section \ref{s:coulomb} with $M_\text{gap}=m$ has 
\be
\label{abarCou}
\text{1-loop Coulomb:}~~~
\bar{a}_{1,0}
=\frac{1}{10} = 0.1\,,~~~
\bar{a}_{2,0} = \frac{1}{70} \approx 0.014\,, ~~~
\bar{a}_{3,0} = \frac{1}{420}\approx 0.0024\,,
~~~
\text{etc.}
\ee
and is shown as a {\bf blue dot}. Since the Coulomb branch 1-loop amplitudes has Wilson coefficients 
$\bar{a}_{k,q}$ that are numerically very small, especially with increasing $k$, we only include the Coulomb point in plots for $k \le 3$. For the same reason, we do not include the curves of the 1-loop amplitudes with $M_\text{gap}/m$ varying between 0 and 1, though they too must lie with the allowed region.

\begin{figure}[t!]
\centering
\includegraphics[width=\textwidth]{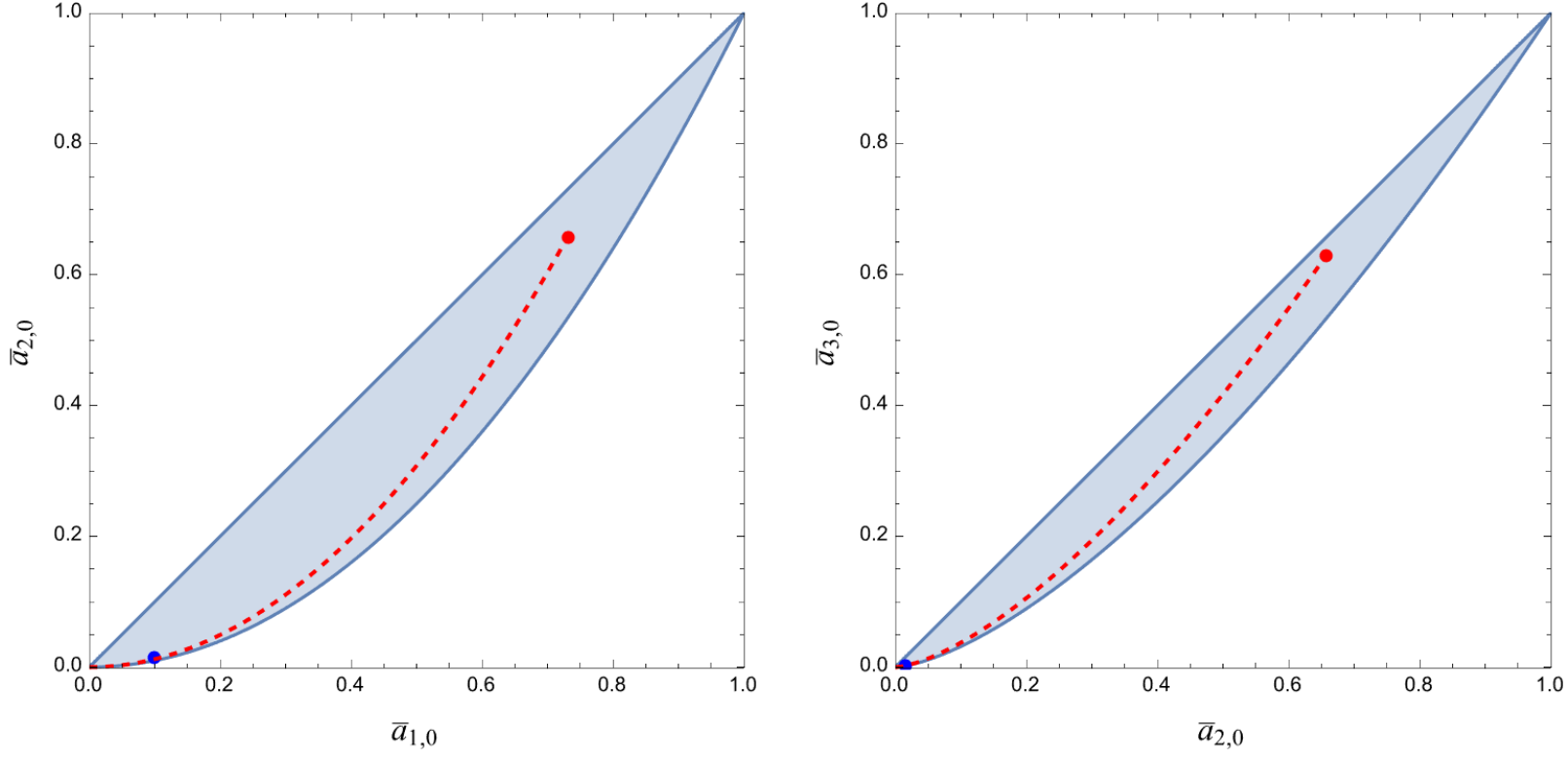}
\caption{\label{fig:fixedprojs} The allowed regions in the $(\bar{a}_{1,0},\bar{a}_{2,0})$ (left) and $(\bar{a}_{2,0},\bar{a}_{3,0})$ (right) planes. The red dots mark the Veneziano amplitude with $M_{\gap}^2 = 1/\a'$ and the dashed line represents the Veneziano amplitude as a function of  $0< \a'M_{\gap}^2 < 1$.
The blue dot the 1-loop Coulomb  amplitude with $M_{\gap}=m$. The lower bounds are $\bar{a}_{2,0} = \bar{a}_{1,0}^{1/2}$  and
$\bar{a}_{3,0} = \bar{a}_{2,0}^{2/3}$, respectively, as discussed in the main text.
}
\end{figure}

\vspace{2mm}
\noindent {\bf $(\bar{a}_{k,0},\bar{a}_{k',0})$ Regions.}
Analytic bounds on the space of Wilson coefficients, such as Hankel matrix and cyclic polytope constraints, were derived in \cite{Arkani-Hamed:2020blm} and extended in \cite{Chiang:2021ziz}. 
In general, for given finite\footnote{It is possible that the bounds would be equivalent in the limit of $k_\text{max},\ell_\text{max}\to \infty$.} $k_\text{max}$ and $\ell_\text{max}$,   
 the collection of these analytic bounds tends to overestimate the allowed regions 
 compared the bounds 
 found with numerical methods such as CPLEX or SDPB. However, in the special case of projections onto the $(\bar{a}_{k,0},\bar{a}_{k',0})$ planes, a finite subset of Hankel constraints imply the region is bounded by
\begin{align}\label{ak0akp0bd}
    \bar{a}_{k,0}^{k'/k} 
\leq \bar{a}_{k',0} \leq \bar{a}_{k,0}
~~~
\text{for $k \leq k'$}\,,
\end{align}
which agrees with the SDPB/CPLEX numerical results. 
The bounds \reef{ak0akp0bd} are independent of $k_{\max}$. 
Figure \ref{fig:fixedprojs} displays the projections into the 
$(\bar{a}_{1,0},\bar{a}_{2,0})$ and $(\bar{a}_{2,0},\bar{a}_{3,0})$ planes as examples of such regions. These plots also show the locations of  the Veneziano amplitude and the 1-loop Coulomb branch within the region.

The infinite spin tower amplitude discussed in Section \ref{s:otherAmps} has Wilson coefficients
\be
 \label{spintower}
 a_{k,q} = \bigg(\frac{M_\text{gap}}{m}\bigg)^{2k+4}
~~~~\implies~~~~
(\bar{a}_{k,0})^{\frac{1}{k}} 
=(\bar{a}_{k',0})^{\frac{1}{k'}}\,.
\ee
This saturates the lower bound on the region \reef{ak0akp0bd}.

Note  that  $M_\text{gap} = m$ corresponds to the point (1,1) in {\em any} 2d projection $(\bar{a}_{k,q},\bar{a}_{k',q'})$, so our 2d plots  always include the (1,1) point. Similarly, the extreme limit $M_\text{gap} \ll m$
corresponds to (0,0) in any such 2d projection; that is the limit of the $\mathcal{N} =4$ SUSY operator $\tr F^4$ having a coupling that  dominates every other operator. 
Because (0,0) and (1,1) are included in all plots,  convexity of the allowed region implies that the diagonal $\bar{a}_{k,q}=\bar{a}_{k',q'}$ is also included. In general it need not correspond to a bound of the region, though it does for the $(\bar{a}_{k,0},\bar{a}_{k',0})$ projections.

\begin{figure}
\centering
\includegraphics[width=\textwidth]{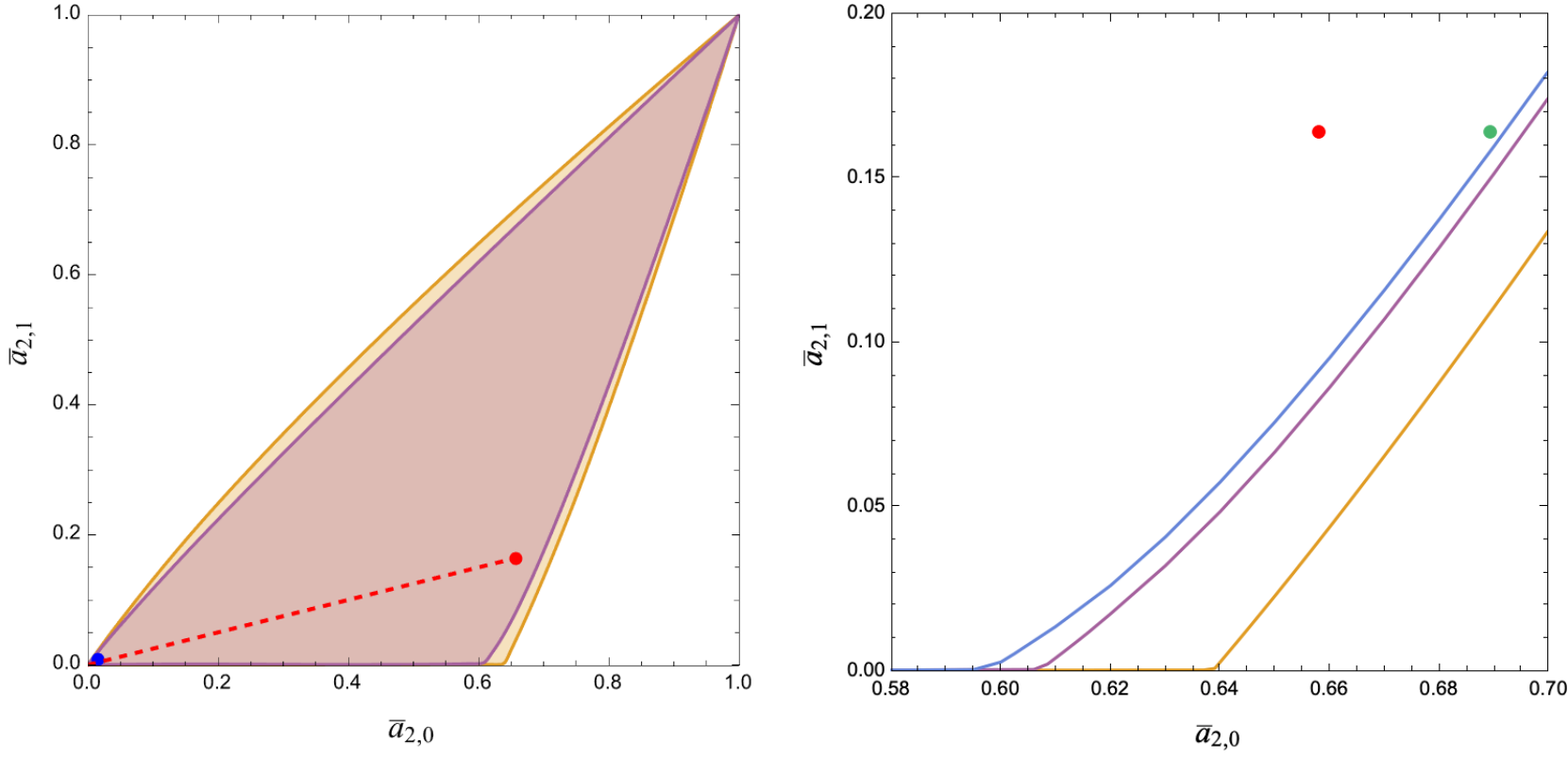}
\caption{\label{fig:a20a21proj}Left: The allowed regions for the projection to the $(\bar{a}_{2,0},\bar{a}_{2,1})$ plane. The orange bounds are for $k_\text{max} = 4$ and $\ell_\text{max} = 200$, while the violet bound is $k_\text{max} = 10$ and $\ell_\text{max} = 300$. (Taking $\ell_\text{max}$ higher results in differences at order $10^{-4}$ or less, not visible in the plot.) The red dot marks the Veneziano amplitude and the blue dot the 1-loop Coulomb amplitude with coefficients \reef{abarVen} and \reef{abarCou}, respectively.
 \newline
Right: Zoom-in on the bounds near the Veneziano amplitude (red) to compare the $k_{\max}=4$ and $10$ bounds with the  $k_{\max} = 15$ bounds obtained with $\ell_{\max} = 800$. The green dot shows the maximum allowed value of $\bar{a}_{2,0}$ for $k_\text{max} = 20$ and $\ell_\text{max} = 600$ when $\bar{a}_{2,1}$ is fixed at the string value. These results give no indication that the bounds converge to the string as $k_\text{max} \to \infty$.}
\end{figure}

\vspace{2mm}
\noindent {\bf The $(\bar{a}_{2,0},\bar{a}_{2,1})$ Region.}
The bounds on the $(\bar{a}_{k,0},\bar{a}_{k',0})$ regions were independent of $k_\text{max}$, but for general projections $(\bar{a}_{k,q},\bar{a}_{k',q'})$ the bounds depend on $k_\text{max}$ and we are interested in how they converge as $k_\text{max} \to \infty$. For that reason, we study the bounds for increasing $k_\text{max}$, with the choice limited only by computation time. 

The simplest example of these $k_{\max}$ dependent regions is the $\bar{a}_{2,1}$ vs. $\bar{a}_{2,0}$ projection, which we display in Figure  \ref{fig:a20a21proj}. The bounds shown were obtained with both SDPB and CPLEX whose results are visually indistinguishable in these plots. A more detailed comparison of the SDPB and CPLEX numerics is presented in Section \ref{s:CPLEXvsSDPB}. Benchmarking for the choices of $\ell_\text{max}$ is discussed in Appendix \ref{app:linprogacc}.

The numerical results indicate that string theory with $\alpha'M_{\gap}^{2} = 1$ is close to, but not on, the boundary of this projection. Moreover, for $k_\text{max} \le 15$, there is no indication of a kink near the string. There does appear to be a kink on the $\bar{a}_{2,0} $-axis, namely where the lower bound on $\bar{a}_{2,1}$ goes from being zero to non-zero. As $k_\text{max}$ increases, the kink moves slowly to lower values of $\bar{a}_{2,0}$; for $k_\text{max}=15$, it is at $\bar{a}_{2,0}$ slightly below $0.6$, but it is not clear what it asymptotes to for $k_\text{max} \to \infty$.\footnote{The amplitude 
\be
A^{\text{NF}}[zz\bar{z}\bar{z}] = -\frac{s}{u}+\frac{s^2}{2M_\text{gap}^2}\left(\frac{1}{M_\text{gap}^2-s}+\frac{1}{M_\text{gap}^2-u}
\right)\,
\label{nonbps0}
\ee
has Wilson coefficients $a_{0,0}=1$, $a_{k,0}=a_{k,k}=1/2$ for $k>0$, and $a_{k,q}=0$ for $0<q<k$. As such, it is a candidate for the point $(1/2,0)$ in any $(\bar{a}_{k,0},\bar{a}_{k',q})$ projection. However, \reef{nonbps0} does not satisfy the Froissart bound \reef{matrinlikeboundss}. (This is similar to the ``spin 1 theory'' discussed in the pion-bootstrap \cite{Albert:2022oes}.) One could speculate that the cusp approaches $(1/2,0)$ in the limit of $k_\text{max} \to \infty$, but at large $k,k'$, the Veneziano amplitude has $(\bar{a}_{k,0},\bar{a}_{k',1}) \to (6/\pi^2,0) \approx (0.608,0)$, so that proposal seems implausible for all $k,k'$. 
}

\begin{figure}
\centering
\begin{tikzpicture}
  \node[anchor=south west,inner sep=0] (image) at (0,0) {\includegraphics[width=\textwidth]{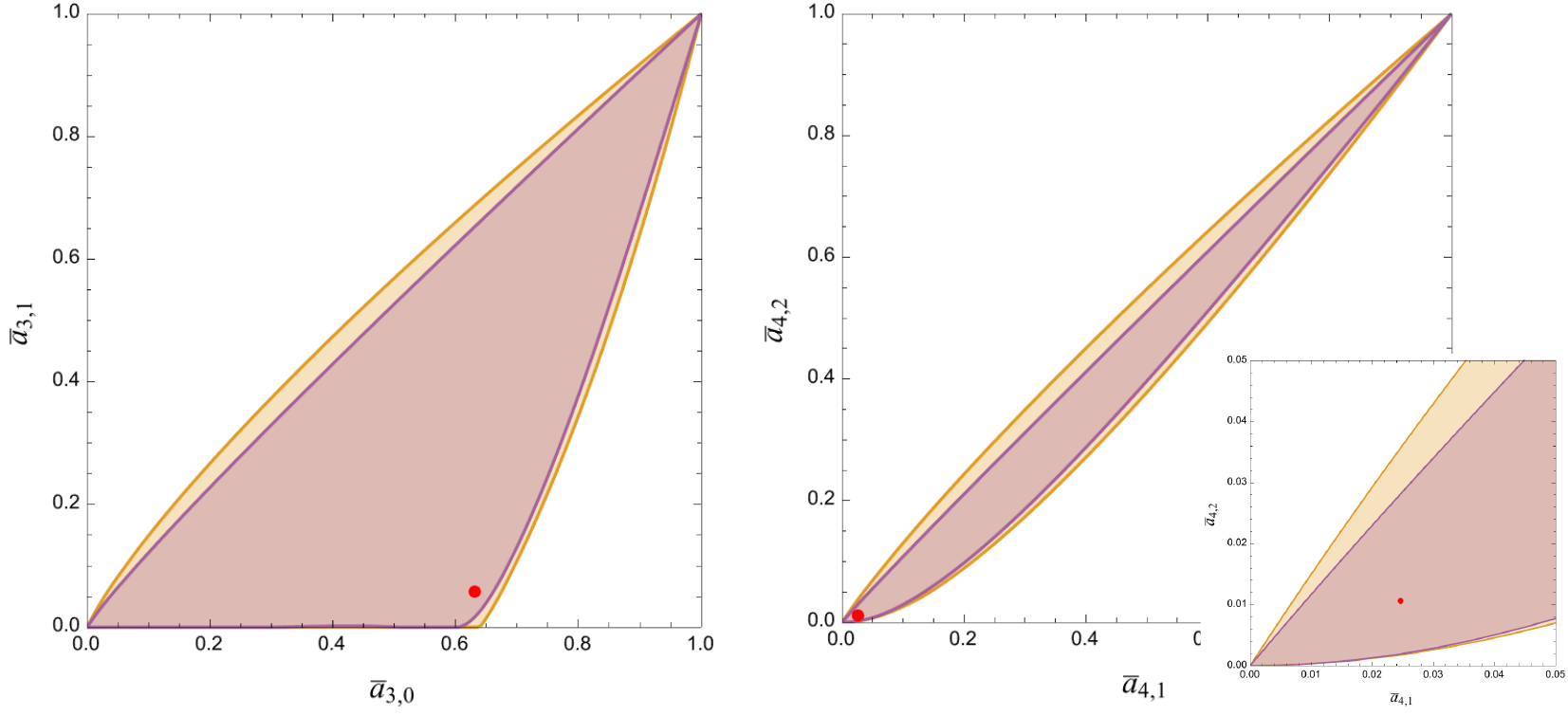}};
  \begin{scope}[x={(image.south east)},y={(image.north west)}]
  \end{scope}
\end{tikzpicture}
\caption{\label{fig:a30a31proj} The allowed regions  in the $(\bar{a}_{3,0},\bar{a}_{3,1})$ and $(\bar{a}_{4,1},\bar{a}_{4,2})$ projections for $k_{\max}=4,\ell_{\max} = 200$ (orange) and $k_{\max}=10,\ell_{\max} = 300$ (purple). The red dot represents the Veneziano amplitude.}
\end{figure}

\vspace{2mm}
\noindent {\bf The $(\bar{a}_{3,0},\bar{a}_{3,1})$ and $(\bar{a}_{4,1},\bar{a}_{4,2})$  
Regions}
We also consider the $\bar{a}_{3,1}$ vs. $\bar{a}_{3,0}$ and $\bar{a}_{4,2}$ vs. $\bar{a}_{4,1}$ projections in Figure \ref{fig:a30a31proj}. 
In both cases, the string again lies close to, but not on, the boundary. The $\bar{a}_{3,1}$ vs. $\bar{a}_{3,0}$ projection is qualitatively similar to the $(\bar{a}_{2,0},\bar{a}_{2,1})$ projection. In particular, it also shows indications of a kink on the horizontal axis, in this case near $\bar{a}_{3,0} \sim 0.6$.

The lower bound in the $(\bar{a}_{4,1},\bar{a}_{4,2})$ projection is qualitatively different from the previous two in that it does not include points on the horizontal axis and there is no indication of a kink. It is noteworthy that the allowed region is very slim: this implies a strong correlation between the allowed coefficients of the corresponding $\mathcal{N}=4$ SUSY $\tr D^8 F^4$ operators.

\begin{figure}
    \centering
    \includegraphics[width=\textwidth]{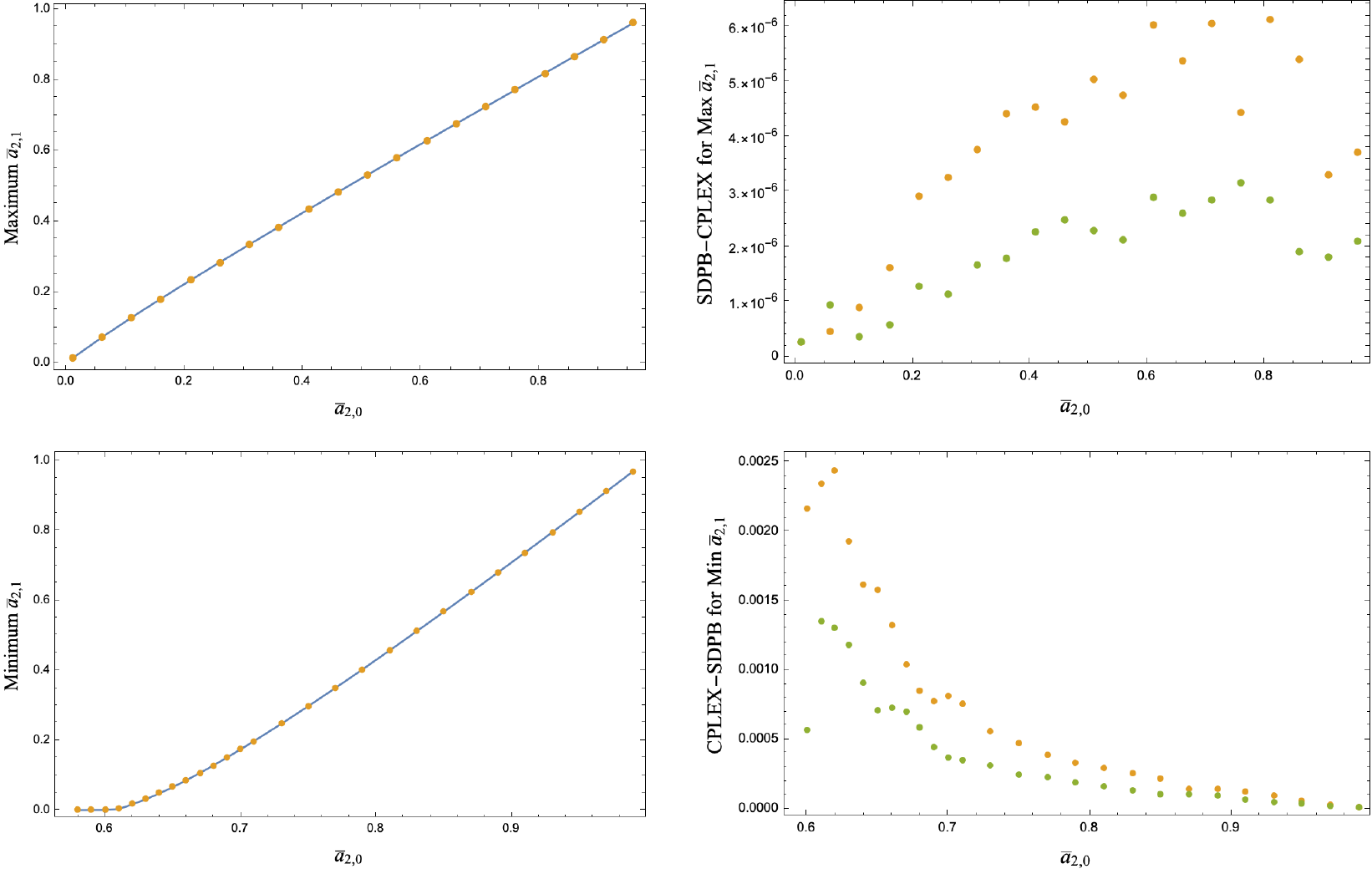}
    \caption{Left: Minimum (top) and maximum (bottom) $\bar{a}_{2,1}$ calculated with SDPB (blue) and CPLEX (orange) at $\ell_{\max} = x_{\max} = 300$. While SDPB is represented as a continuous curve, the code is run at the same set of points as CPLEX and the points are then joined so they can be distinguished from the CPLEX results. \newline
    Right: The absolute difference between SDPB and CPLEX for the points given on the left for $x_{\max} = 300$ (orange) and $x_{\max} = 500$ (green). As expected, SDPB gives a slightly larger allowed region because it does not rely on discretizing $x$, and the agreement becomes better as we increase $x_{\max}$.}
    \label{fig:a20SDPBvsCPLEX}
\end{figure}

\subsection{Comparison of SDPB and CPLEX}
\label{s:CPLEXvsSDPB}

Computing the bounds in both SDPB and CPLEX provides a cross-check on the numerical methods. We find excellent agreement between these techniques.

As an example, the bounds in Figure \ref{fig:a20a21proj} were computed with both SDPB and CPLEX. Figure  \ref{fig:a20SDPBvsCPLEX} shows the difference between the upper and lower bounds for $k_{\max} = 10$ and $\ell_{\max}=300$ as obtained by both methods, using $x_{\max} =300$ for CPLEX.

\begin{figure}
    \centering
    \includegraphics[width=\textwidth]{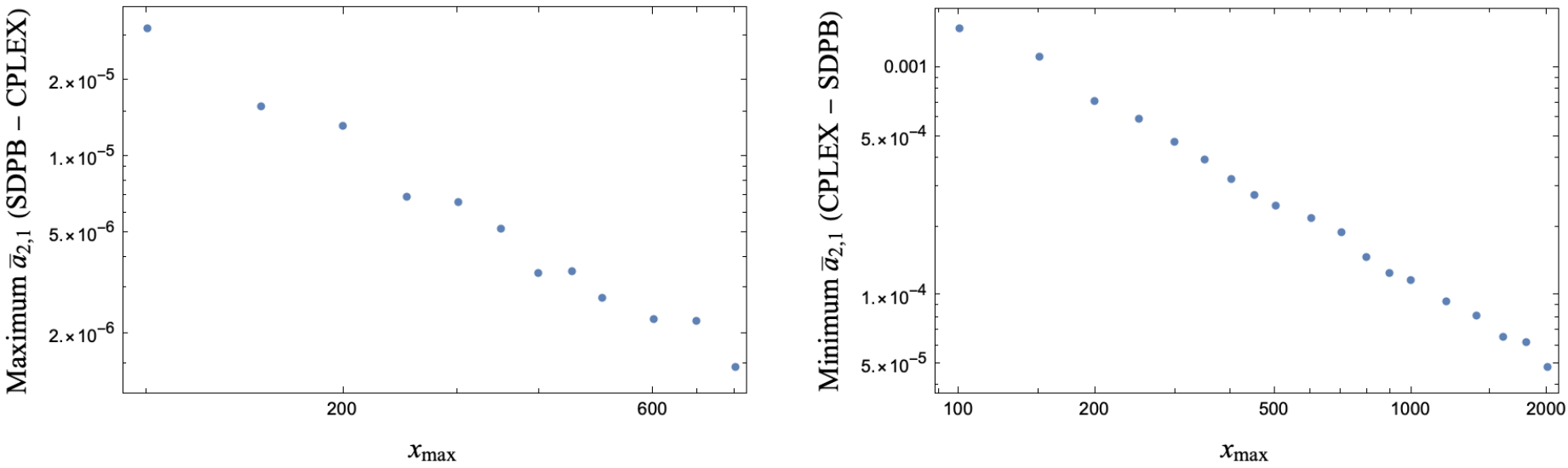}
    \caption{We show the convergence of the CPLEX bounds to the SDPB bounds goes as a power law in $x_{\max}$ for both the maximum (left) and minimum (right) $\bar{a}_{2,1}$ with $\bar{a}_{2,0} = 3/4$ fixed and $\ell_{\max} = 300$.}
    \label{fig:xmaxconvg}
\end{figure}

Because of the discretization, CPLEX underestimates the allowed space slightly compared to SDPB, but the difference becomes increasingly small with increasing discretization parameter $x_\text{max}$.
This is illustrated in 
Figure  \ref{fig:xmaxconvg} which shows that the CPLEX bounds converge to the SDPB result as a power law in $x_{\max}$.

In terms of computation time, CPLEX with lower values of $x_\text{max}$ runs faster than SDBP. However, for high precision results, higher $x_\text{max}$ is needed and the time-advantage goes away. For high-precision, we find SDPB faster and more reliable. 
In the remainder of the paper, all plots are made with SDPB  while CPLEX is used for basic ``sanity-checks''.

\section{Veneziano from String Monodromy}\label{sted}

\subsection{String Monodromy} 
The tree-level amplitudes in Type-I string theory can be written as  period integrals multiplied by a universal pre-factor. Specifically at 4-point, we have 
\begin{equation}\label{stringsol}
\begin{split}
A[z_1 z_2 \bar{z}_3 \bar{z}_4]
&=-\frac{\alpha's^{2}}{t} \,\int_{0}^{1}dz z^{-\alpha's-1}(1-z)^{-\alpha'u-1} \ , \\ 
A[z_1 \bar{z}_3 z_2\bar{z}_4]
&=\frac{\alpha's^{2}}{t}\, \int_{1}^{\infty}dz z^{-\alpha's-1}(z-1)^{-\alpha'u-1}  \ ,\\ 
A[z_2 z_1 \bar{z}_3 \bar{z}_4]&=\frac{\alpha's^{2}}{t}\, \int_{-\infty}^{0}dz (-z)^{-\alpha's-1}(1-z)^{-\alpha'u-1} \ .
\end{split}
\end{equation}
Here and below, 
$\bar{z}$ are the complex $\mathcal{N}=4$ scalars introduced in Section \ref{s:superA}. 
 $A[z_1 z_2 \bar{z}_3 \bar{z}_4]$ is the Veneziano amplitude \reef{veneziano} and the two other amplitudes are the color rearranged versions of it.

The three amplitudes \reef{stringsol} differ only by their integration region. A contour deformation \cite{Plahte:1970wy,Stieberger:2009hq,Bjerrum-Bohr:2009ulz,Bjerrum-Bohr:2010mia,Bjerrum-Bohr:2010pnr}
relates the three amplitudes linearly to each other, with monodromy factors picked up at $z=0$ and $z=1$. The resulting  
 4-point {\em string monodromy relation} is 
\begin{equation}\label{monodromrel}
0 = A[z_2 z_1 \bar{z}_3 \bar{z}_4]+e^{i\pi\a's}A[z_1 z_2 \bar{z}_3 \bar{z}_4]+e^{-i\pi\a't}
A[z_1 \bar{z}_3 z_2\bar{z}_4]
\end{equation}
Now using the SUSY Ward identities
\reef{twoA4s} and that $A[z_1z_2\bar{z}_3\bar{z}_4] = s^2 f(s,u)$, where $f$ is real, we can write the real and imaginary parts of \reef{monodromrel} as
\be
\label{monodromrelReIm}
\begin{split}
    0 &= f(s,t)+\cos(\pi\a's)f(s,u) + \cos(\pi\a't)f(t,u)\,,
    \\[2mm]
    0 &= \sin(\pi\a's)f(s,u)-\sin(\pi\a't)f(t,u)\,.
\end{split}
\ee

Let us impose the monodromy relation on the low-energy expansion 
of the $\mathcal{N} = 4$ SUSY EFT. We plug in the low-energy ansatz 
\reef{fsu}, along with the SUSY crossing constraints \reef{crossingsymmet}, and solve \reef{monodromrelReIm} order by order in the Mandelstam expansion. This fixes particular linear combinations of Wilson coefficients as shown in Table \ref{monoconk4}. There and in the remainder of this section we set $\alpha'=1$ and $M_\text{gap}=1$. 
\begin{table}[t]
\be
\nonumber
\begin{array}{llc}
\text{linear combination fixed} &
\text{string value}
& \text{monovariable}
\\
\hline\\[-3mm]

\!\!\raisebox{-1.6mm}{
 $a_{0,0}$
 } 
 & 
\!\!\raisebox{-1.6mm}{
 $=\zeta_{2} =\frac{\pi^{2}}{6}$
  }
  &r_{0}^{(0)}
\\[2.9mm]
a_{2,0} &= \zeta_{4} =\frac{\pi^{4}}{90}   
&r_1^{(2)}
\\[2.5mm]
a_{2,1} &= 
\frac{1}{4}\zeta_{4}
=\frac{\pi^{4}}{360}    &r_2^{(2)}
\\[2.5mm]
a_{3,1}-2a_{3,0}+\zeta_2\, a_{1,0} &= 0  &r_3^{(3)}
\\[2.5mm]
a_{4,0} &= \zeta_6 = \frac{\pi^{6}}{945}  &r_4^{(4)}
\\[2.5mm]
a_{4,2}-2a_{4,1} &= -\frac{1}{16} \zeta_6
=
-\frac{\pi^{6}}{15120}~ &r_5^{(4)}
\\[2.5mm]
a_{5,1} - 3 a_{5,0} + \zeta_2 a_{3,0} + \zeta_4 a_{1,0} &=0
&r_6^{(5)}
\\[2.5mm]
a_{5,2} - 5 a_{5,0} + 2\zeta_2\, a_{3,0} + \frac{5}{4}\zeta_4 \,a_{1,0} &=0
&r_7^{(5)}
\end{array}
\ee
\caption{\label{monoconk4}The string monodromy relation \reef{monodromrel} fixes particular linear combination of the Wilson coefficients $a_{k,q}$ in the supersymmetric ansatz \reef{ansatz}-\reef{crossingsymmet} as shown here up to $k=5$ with $\alpha'=1$. The monovariables were introduced in Section \ref{s:intro} and are reviewed in Section \ref{s:flat}.}
\end{table} 

The monodromy relations do not fix all Wilson coefficients. The Wilson coefficients {\em unfixed} by monodromy with $k\leq 8$ are 
\be 
\label{aunfixed}
a_{1,0}\,,~~ 
a_{3,0}\,,~~ 
a_{4,1}\,,~~
a_{5,0}\,,~~
a_{6,1}\,,~~
a_{7,0}\,,~~
a_{7,2}\,,~~
a_{8,1}\,.
\ee
 Comparing to the Veneziano amplitude (with $\alpha'=1$), these monodromy-unfixed coefficients  all involve $\zeta_\text{odd}$: we have
\begin{equation}
\label{astrvalues}
\begin{split}
&a_{1,0}^{\text{str}}=\zeta_{3}\,,~~
a_{3,0}^{\text{str}}=\zeta_{5}\,, ~~
a_{4,1}^{\text{str}}= \tfrac{3}{4} \zeta_6 -\tfrac{1}{2}
\zeta_3^2\,,~~
a_{5,0}^{\text{str}}=\zeta_{7} \,, ~~
a_{6,1}^{\text{str}} = \tfrac{5}{4} \zeta_8 -
\zeta_3 \zeta_5\,,\\[2mm]
&
a_{7,0}^{\text{str}}=\zeta_{9} \,,~~
a_{7,2}^{\text{str}}=
-\tfrac{7}{4} 
\zeta_6 \zeta_3 
+ \tfrac{1}{6} \zeta_3^3 - \tfrac{9}{4} \zeta_4 \zeta_5 
- 3 \zeta_2 \zeta_7 
+ \tfrac{28}{3} \zeta_9\,,
~~
a_{8,1}^{\text{str}}=
\tfrac{7}{4}  \zeta_{10} - \tfrac{1}{2} \zeta_5^2 - \zeta_3 \zeta_7 \,.
\end{split}
\end{equation}
The monodromy relations  only ``know'' $\pi$, i.e.~$\zeta_\text{even}$, so they cannot fix the $\zeta_\text{odd}$-dependence in the amplitude.

\subsection{Bootstrapping Veneziano}
\label{sec:monoconj}

Huang, Liu, Rodina, and Wang \cite{Huang:2020nqy} found numerical evidence that when a subset of analytic EFT-hedron bounds from \cite{Arkani-Hamed:2020blm} were combined with the monodromy constraints, $a_{1,0}$, $a_{3,0}$, and $a_{4,1}$
were within 1.5\%, 0.2\%, and 53\% of the string values \reef{astrvalues}.\footnote{The new paper \cite{yutinetal} improves these bounds.}

\begin{figure}[t]
\centering
\includegraphics[width=\textwidth]{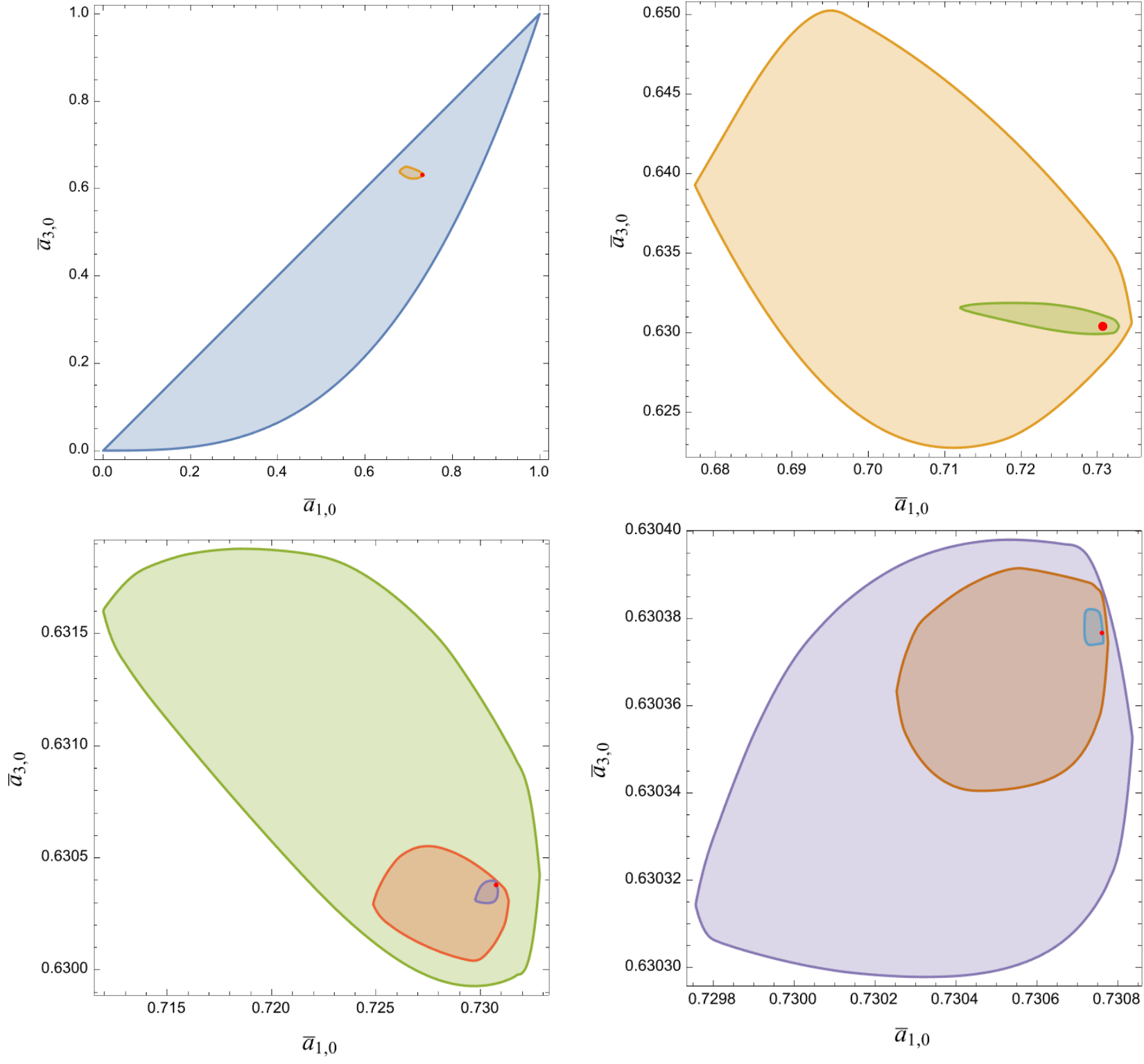}
\caption{\label{fig:a30a10proj} Regions allowed by SDPB bounds on the $\bar{a}_{1,0}$ vs.~$\bar{a}_{3,0}$ when monodromy and crossing are imposed up to a given $k_{\max}$ with $\ell_{\max} = 800$. The blue region on the top left is the exact allowed region without monodromies imposed. The red dot marks the Veneziano amplitude.}
\end{figure}

To extend the results of \cite{Huang:2020nqy}, we include the monodromy constraints in Table  \ref{monoconk4} as additional null constraints in the formulation of the linearized optimization problem in \reef{vectorequation}-\reef{vectorequation2} and work systematically up to $k_\text{max} = 8$.

Starting with the $(\bar{a}_{1,0},\bar{a}_{3,0})$ region, we know from Section \ref{s:exbounds} that {\em without} the monodromy constraints, the allowed region is bounded 
as $\bar{a}_{1,0}^3 \le \bar{a}_{3,0} \le \bar{a}_{1,0}$. This is the blue region in the top-left plot in Figure \ref{fig:a30a10proj}. In that same plot, the orange region is the allowed region found with SDPB when monodromy constraints are imposed to order $k_\text{max} = 3$. The red dot within the $k_\text{max} = 3$ monodromy region is the Veneziano amplitude. 
Zooming in on the orange $k_\text{max} = 3$ monodromy region, we increase $k_\text{max}$ up to 8, as progressively shown in the three other plots in Figure \ref{fig:a30a10proj}, to see how a smaller and smaller island around the Veneziano amplitude is isolated. This progression indicates that the intersection of the monodromy plane and the allowed supersymmetric EFT-hedron region shrinks to a point in the limit $k_\text{max} \to \infty$ as anticipated by the authors of \cite{Huang:2020nqy}.

\begin{figure}[t]
\centering
\includegraphics[width=\textwidth]{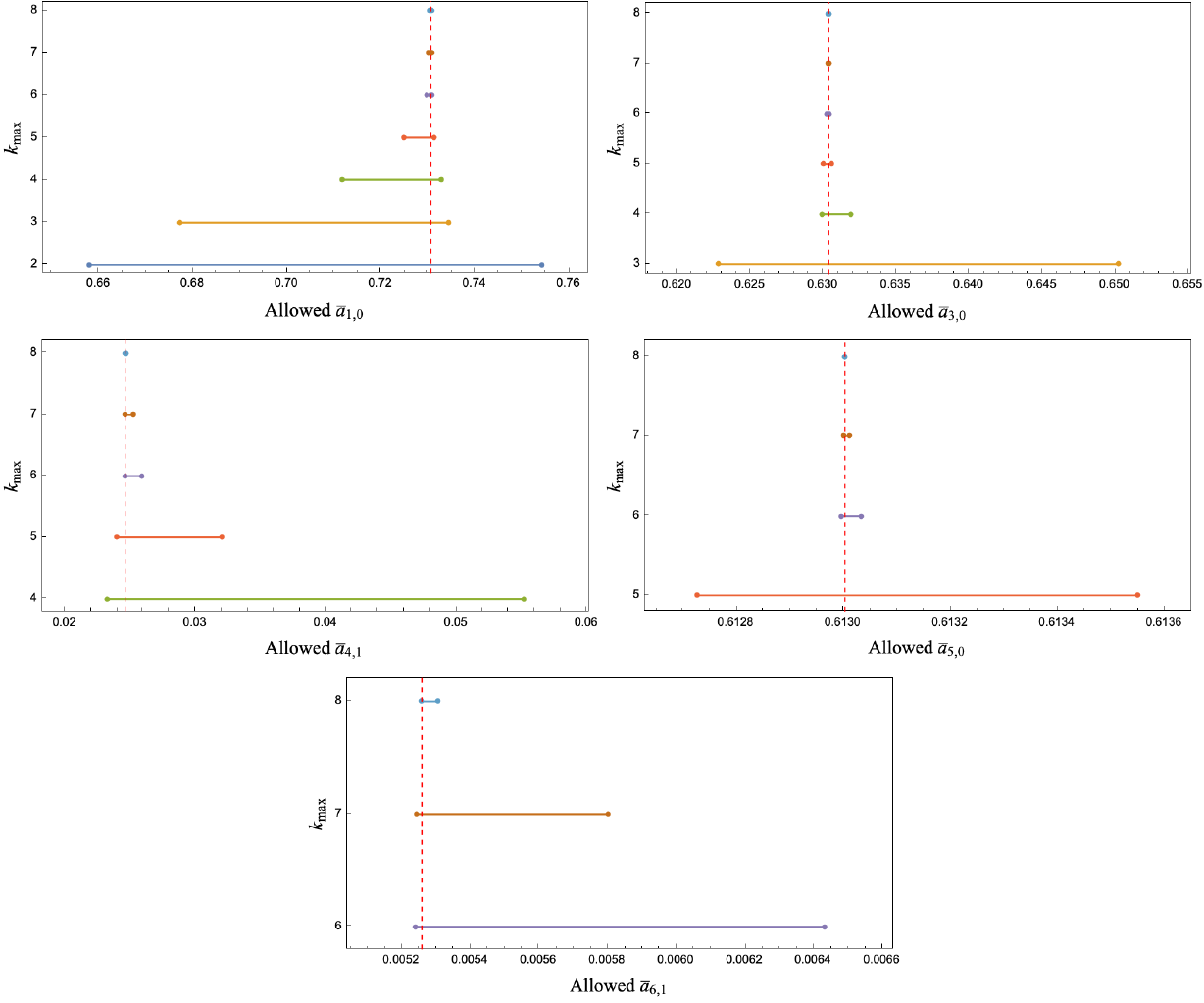}
\caption[monoconj]{SDPB bounds with the monodromy imposed. As $k_{\max}$ is increased, the allowed range of each Wilson coefficient shrinks.
These bounds were found with $\ell_{\max} = 500$ for $k_{\max} = 2,3,\ldots,7$ and $\ell_{\max} = 800$ for $k_{\max} = 8$.}
\label{fig:monoconj}
\end{figure}

A similar result is found for the other coefficients \reef{aunfixed} that were unrestricted by monodromy. At $k_\text{max} = 8$, the bounds we find (working at $\ell_\text{max} = 800$) are 
\begin{equation}
\begin{array}{rcll}
\text{{\bf SDPB}}
&\!\!\text{{\bf bounds}}
&~~&
\text{{\bf String Value}}
\\
1.201982&\leq a_{1,0}\leq &1.202061 \quad & 1.202057 \\
1.036923&\leq a_{3,0}\leq& 1.036937 \quad 
&1.036928 \\
0.04053 &\leq a_{4,1}\leq& 0.04063 \quad & 0.04054 \\
1.0083481 &\leq a_{5,0}\leq &1.0083495 \quad & 1.0083493 \\
0.008649 &\leq a_{6,1}\leq &0.008729 \quad &0.008651 \\
1.00200830 &\leq a_{7,0}\leq &1.00200891 \quad & 1.00200839 \\
0.00031 &\leq a_{7,2}\leq &0.00041 \quad &0.00032 \\
0.00203 &\leq a_{8,1}\leq &0.00212 \quad &0.00204 \\
\end{array}
\end{equation}
Our bounds bring $a_{1,0}$, $a_{3,0}$, and $a_{4,1}$ within 0.0066\%, 0.0013\%, and 0.24\% of the string value. 
The shrinking of the allowed ranges with increasing $k_\text{max}$ is visualized for the first five coefficients in Figure \ref{fig:monoconj}.

\section{Flattening of the EFT-hedron}
\label{s:flat}

In the previous section, we provided new numerical evidence that the supersymmetric EFT-hedron constraints together with monodromy select an island that shrinks around the open string Veneziano amplitude, 
as first proposed in \cite{Huang:2020nqy}. In this section, we explore the geometric consequences of this phenomenon and present evidence for flattening of the allowed space. We also reparameterize the low-energy expansion of the amplitude to coefficients motivated by the flattening and show that it can be partially resummed. 

\subsection{Flattening  Conjecture}

From a geometric perspective, we can think of the linear monodromy constraints, listed at lowest orders in Table \ref{monoconk4}, as defining a higher-dimensional plane in the
space of Wilson coefficients. We call this the ``monodromy plane''. The claim of \cite{Huang:2020nqy} is that the monodromy plane and the supersymmetric EFT-hedron intersect each other at a point in the limit $k_\text{max} \to \infty$ and that this point corresponds to the Veneziano amplitude.

We discussed in the Introduction the two different ways the intersection may happen: as illustrated in 
Figure \ref{fig:cartoon}, either the monodromy plane is tangent to the EFT-hedron or the EFT-hedron must
flatten in such a way that the intersection with the monodromy plane shrinks to a point. To assess which option is realized,
imagine taking the monodromy plane in Figure \ref{fig:cartoon} and shifting it around. If the monodromy plane were tangent to SUSY EFT-hedron, some shifts would give no solution at all while others would result in convergence to a finite size region of parameters, unlike the continued shrinking towards a point. On the other hand, if the EFT-hedron itself is flattening, then the shifted monodromy plane should continue to intersect the space at a single point. 
In Section \ref{sec:emcodconj2}, we
vary the monodromy plane in a controlled way and find evidence that the latter option is realized: the EFT-hedron becomes increasingly narrow as $k_{\max}$ gets larger.

The result is that the supersymmetric EFT-hedron must be flattening in certain directions when $k_\text{max}$ increases. 
Specifically, at large $k_\text{max}$, the number of independent Wilson coefficients increases as $\mathcal{O}(k_\text{max}^2/4)$ after imposing the SUSY crossing constraints. Hence, the ``naive'' dimension of the SUSY EFT-hedron is $\mathcal{O}(k_\text{max}^2/4)$ at large $k_\text{max}$. The monodromy relations fix linear relations among
$\mathcal{O}(k_\text{max}^2/6)$ of these coefficients, thus leaving $1/3$ of the Wilson coefficients unfixed. 

\subsection{Evidence for Flattening} \label{sec:emcodconj2}

The monodromy relations fix certain linear combinations of the Wilson coefficients to particular values. If we simply change those values, we can move the monodromy plane in a controlled way. To do so, define the linear combinations fixed by monodromies to be ``monovariables'' $r_i^{(k)}$, where $k$ denotes the largest $k$ value for any $a_{k,q}$ that appears in the linear combination.
It follows from Table \ref{monoconk4} that:
\be
\label{k4mono}
  r_{0}^{(0)} =a_{0,0}\,,
  ~~~
  r_{1}^{(2)}=a_{2,0}\,,
  ~~~
  r_{2}^{(2)}=a_{2,1}\,,
  ~~~
  r_{3}^{(3)}
=a_{3,1}-2a_{3,0}+\zeta_2\,a_{1,0}\,,
\ldots
\ee
Note that we use the linear combination of the monodromy relations with $\alpha'=1$ for simplicity. We could have reintroduced the scale as $M_\text{gap}$ or another mass $m$.
For string theory, the monovariables take on the values shown in  Table \ref{monoconk4}. Changing the values changes the underlying theory and means the new constraints may no longer necessarily correspond to some relationship between color-ordered amplitudes.

One way to systematically generate new examples of monovariables is to exploit convexity of the SUSY EFT-hedron and use linear combinations of known models to construct more general points inside the allowed space. This way, it is also known what the values of remaining unfixed $a_{k,q}$'s are so there is a check on the numerical bootstrap. 

\begin{figure}
\centering
\includegraphics[width=0.9\textwidth]{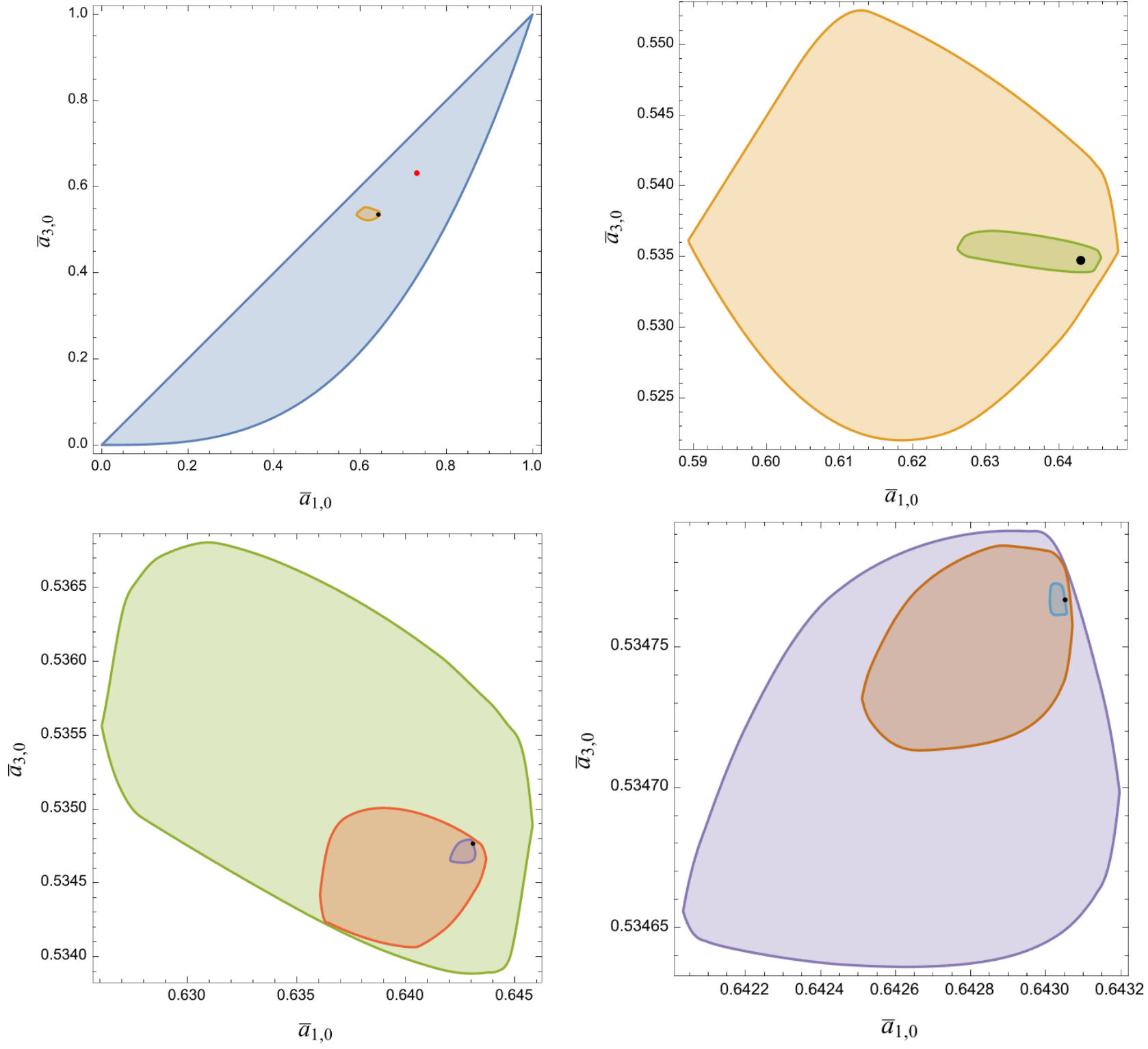}
\caption{\label{fig:a30a10projrt}Plots of the allowed $(\bar{a}_{1,0},\bar{a}_{3,0})$ region when $r_{i}^{(k)}/r_{0}$ variables and crossing are imposed up to a given $k_{\max} = 8$ and $\ell_{\max} = 800$ for the test example  specified in equation \reef{exModel}. The Veneziano amplitude is indicated with a red dot, whereas the test model is shown as a black dot.   Qualitatively this is very similar to the string monodromy case in Figure \ref{fig:a30a10proj}, but quantitatively the islands isolate a different point in the SUSY EFT-hedron.}
\end{figure}

To test the flattening of the SUSY EFT-hedron, we consider linear combinations of the infinite spin tower amplitude, the Veneziano amplitude, and the one-loop Coulomb branch amplitude. To be specific, we consider an ansatz of the form 
\begin{equation}\label{ansatzzz}
\begin{split}
A[zz\bar{z}\bar{z}]=\frac{-s}{u}+s^{2}\Bigg(&\int_{1}^{\infty} dm^{2} \,\rho_{m^{2}}^{(1)}\left [ \frac{1}{(m^{2}-s)(m^{2}-u)} \right ] \\
&+\int_{1}^{\infty} dm^{2} \,\frac{\rho_{m^{2}}^{(2)}}{m^{4}}\left [\frac{m^{4}}{su}- \frac{\Gamma(-s/m^{2})\Gamma(-u/m^{2})}{\Gamma(1+t/m^{2})}\right ] \\
&+\int_{1}^{\infty} dm^{2}\, \frac{\rho_{m^{2}}^{(3)}}{m^4} \,F_3\Big( 1,1,1,1;\frac{5}{2} \Big| \frac{s}{4m^2},\frac{u}{4m^2}\Big)\Bigg)\,,
\end{split}  
\end{equation}
where we work with $M_{\gap}^{2}=1$. The ansatz in  (\ref{ansatzzz}) obeys crossing symmetry by construction. Positivity $\rho_{m^{2}}^{(I)} \geq 0$ is ensured by the densities being randomized positive sums over $\d$-functions:
\begin{equation}
\rho^{(I)}_{m^2} = \sum_{i}a_{i}^{(I)}\,
\delta\big(m^{2}-m_{(I),i}^{2}\big) \,
~~~\text{with}~~~
a_{i}^{(I)} \ge 0\,.
\end{equation}

As an example of the procedure, let us choose the values 
\begin{equation}
\label{exModel}
\hspace{-2cm}\text{Test Example:}
~~~~~
\begin{array}{c|c c c| c c c}
     \rho_{m^2}^{(I)} & m^2_{(I),1} & m^2_{(I),2} & m^2_{(I),3} & a^{(I)}_1 & a^{(I)}_2  & a^{(I)}_3  \\
     \hline
     \rho_{m^2}^{(1)} & 7 & 19 & 21 & \frac{80}{53} & \frac{98}{57} & \frac{81}{23}\\
     \rho_{m^2}^{(2)} & 1 & 4 & 28 & \frac{90}{101\zeta_2} & \frac{9}{5\zeta_2} & \frac{63}{19\zeta_2} \\
     \rho_{m^2}^{(3)} & 3 & 15 & 23 & \frac{1}{103} & \frac{2}{77} & \frac{4}{91} \\
\end{array}
\end{equation}

Expanding \reef{ansatzzz} with these choices for $\rho_{m^2}^{(I)}$ gives the Wilson coefficients\footnote{In contrast, the numerical string values \reef{akq-veneziano} are
\be
\nonumber
a^\text{str}_{0,0} = 1.64493\,, 
~~
a^\text{str}_{1,0} = 1.20206\,, 
~~
a^\text{str}_{2,0} = 1.08232\,, 
~~
a^\text{str}_{2,1}  = 0.270581\,.
\ee
so the test model is in a  different part of parameter space.}
\begin{align}
\label{exWCs}
    a_{0,0} = 1.05265\,, 
    & & a_{1,0} = 0.676907\,,& & a_{2,0} = 0.591605\,,& & a_{2,1}  = 0.148397\,, \,~\ldots
\end{align}
which
 lead to monovariable values
\begin{align}\label{monoex}
&\frac{r_1^{(2)}}{r_0^{(0)}} = 0.562015\,, & 
&\frac{r_2^{(2)}}{r_0^{(0)}} = 0.140974\,, & &\frac{r_3^{(3)}}{r_0^{(0)}} = 0.038116\,,  & &\frac{r_4^{(4)}}{r_0^{(0)}} = 0.523818\,, ~\ldots
\end{align}

 Imposing  (\ref{monoex}) as null constraints along with the $\mathcal{X}$ and $\mathcal{Y}$ crossing constraints \reef{crossingsymmimpy}-\reef{crossingsymmimpy2} isolates islands of the allowed space that decrease in size as we increase $k_{\max}$, as shown in Figure \ref{fig:a30a10projrt}, just like the case when the actual monodromy relations isolate an island around the string in Figure \ref{fig:a30a10proj}. 
 We use SDPB to fix the non-monovariables to the ranges ($k_{\max}=8$):
\begin{equation}
\begin{array}{rcll}
\text{{\bf SDPB}} 
& \text{{\bf bounds}}
&  \phantom{0.676912} \quad  \quad  \quad  & 
\text{{\bf Model Value}}
\\
0.676864& < a_{1,0} < &0.676913 & 0.676907\\
0.562915 &< a_{3,0} <& 0.562928 
& 0.562921\\
0.021981& < a_{4,1} <& 0.022031 \quad
& 0.021983\\
0.5463085 &< a_{5,0} < &0.5463095 
&0.5463094\\
0.004685& < a_{6,1} <
& 0.004729 
&0.004687\\
0.5428093& < a_{7,0} < &0.5428099
&0.5428094\\
0.00017274 &< a_{7,2} <& 0.00022540
&0.00017362\\
0.0011034 &< a_{8,1} < &0.0011495
& 0.0011039 \,.
\end{array}
\end{equation}
Here ``Model Value'' is the value of the coefficient for the theory we constructed in \reef{exModel}. For the first three cases, SDBP gets within $0.007\%$, 
$0.002\%$, and $0.22\%$, respectively, of the known model value.

\begin{figure}[t]
\centering
\includegraphics[width=0.6\textwidth]{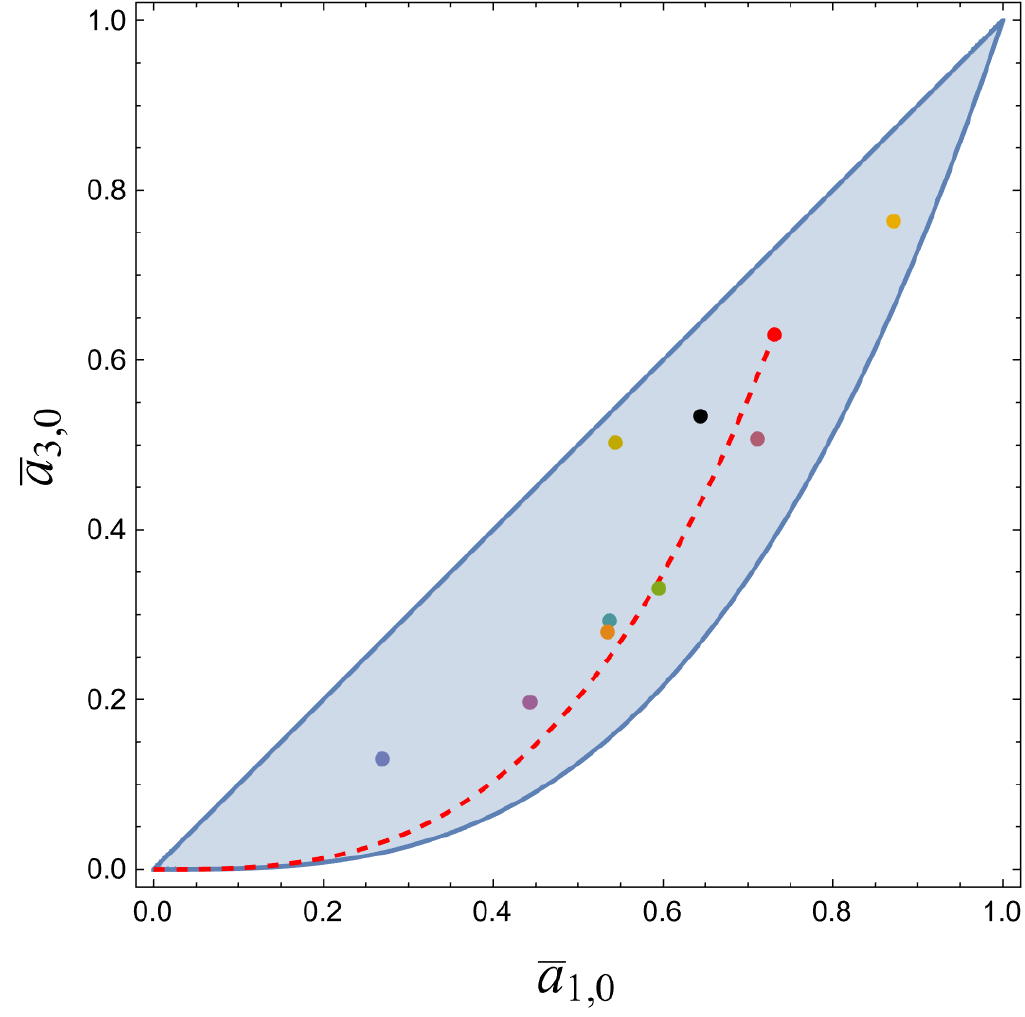}
\caption[rthylocs]{Locations of a selection of test models in the $(\bar{a}_{1,0},\bar{a}_{3,0})$ plane. The colors match those in Figure \ref{fig:emercodimconj}. The red dot marks the Veneziano amplitude.}
\label{fig:rthylocs}
\end{figure}

\begin{figure}[t]
\centering
\includegraphics[width=\textwidth]{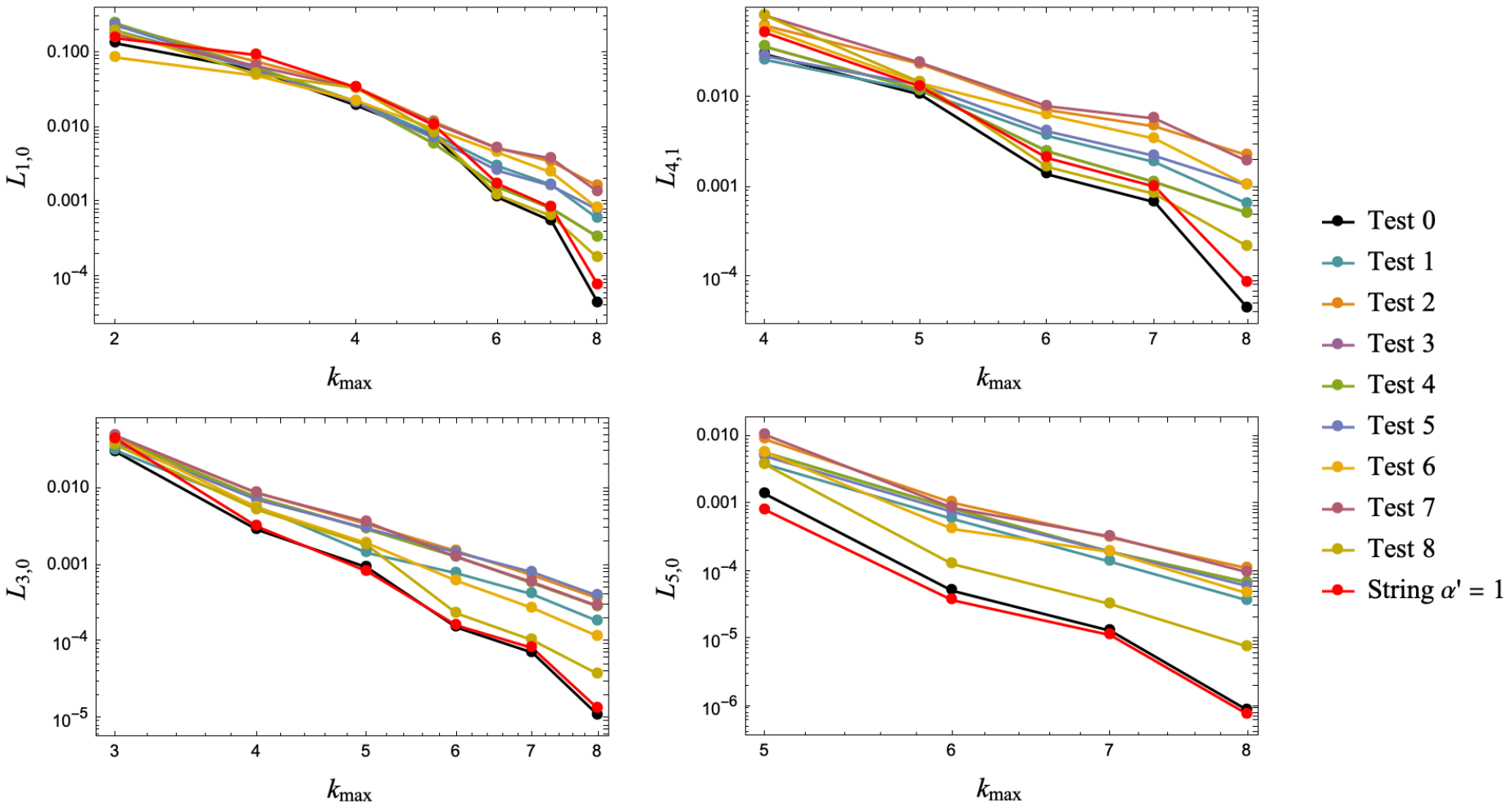}
\caption[emercodimconj]{Interval lengths $L_{k,q}$, as defined in \reef{Ldefined}, as a function of $k_\text{max}$ for a sample of models of the form \reef{ansatzzz} found with SDPB  at $\ell_{\max} = 500$. 
``Test 0'' corresponds to the example test model \reef{exModel}, while the other test theories are created from randomly generated values for $a_{i}^{(I)}$ and $m_{(I),i}^2$. It is illustrated in Figure \ref{fig:rthylocs} where these models lie in the $(\bar{a}_{1,0},\bar{a}_{3,0})$ plane.}
\label{fig:emercodimconj}
\end{figure}

We have run multiple other test theories for which the $\rho_{m^{2}}^{(I)}$ are chosen to be random positive sums of hundreds of different delta functions with a randomly generated mass spectrum above $M_\text{gap}=1$. 
A sample of these test theories is shown in Figure \ref{fig:rthylocs} to illustrate how varying the monovariables allow us to intersect the $(\bar{a}_{1,0},\bar{a}_{3,0})$ plane in widely different locations. 
Remarkably, we find a similar behavior as above for each of the test theories: 
the SDPB bounds narrow in on the known values for each of the Wilson coefficients left unfixed by the monovariable constraints. To illustrate this, consider the interval lengths of the SPDB bounds
\be
  \label{Ldefined}
  L_{k,q} = (a_{k,q})_\text{max}-(a_{k,q})_\text{min}\,.
\ee
For a sample of test models, Figure \ref{fig:emercodimconj} shows  how the $L_{k,q}$'s for the $a_{1,0}$, $a_{3,0}$, $a_{4,1}$, and $a_{5,0}$ tend to zero as $k_\text{max}$ is increased. For comparison, the string is shown in red. These log-log plots indicate that 
each $L_{k,q} \to 0$ at least as a power law in $k_\text{max}$.

\subsection{Good EFT-hedron ``Coordinates''}
\label{s:goodcoords}
The flattening of the allowed space shows that there are stronger constraints among certain combinations of Wilson coefficients than one would naively have expected. This suggests that there is a different low-energy expansion that makes these correlations more manifest. 

To work towards such an alternate representation of the amplitude, 
we start with the general low-energy ansatz \reef{ansatz} and use the monovariable definitions in Table \ref{monoconk4} to bring the monovariables  $r^{(k)}_i$ directly into the parameterization of the amplitude. Thus, in \reef{ansatz}, we replace 
\be
\begin{split}
&
a_{0,0} \to r^{(0)}_0\,,
~~~
a_{2,0} \to r_1^{(2)}\,,
~~~
a_{2,1} \to r_2^{(2)}\,,
~~~
a_{3,1} \to  2a_{3,0}-\zeta_2a_{1,0}+r_3^{(3)}\,, \\
&
a_{4,0} \to r_4^{(4)}\,,
~~~
a_{4,2} \to 2a_{4,1}  +r_5^{(4)}\,,
~~~
a_{5,1} \to  3 a_{5,0} - \zeta_2 a_{3,0} - \zeta_4 a_{1,0} + r_6^{(5)}\,,
~~~\text{etc.}
\end{split}
\ee
We organize the terms in the amplitude into two groups: those with monovariable coefficients 
$r_i^{(k)}$, which each multiply a simple degree $k$ polynomial symmetric in $s$ and $u$, and those with the remaining $a_{k,q}$ variables which each multiply an infinite tower of $s$-$u$ symmetric polynomials  starting at degree $k$. Specifically, we find
\be \label{newexpA4again}
  A[zz\bar{z}\bar{z}] = - \frac{s}{u}
  + s^2
  \bigg( \sum_{k,i}  r_i^{(k)} P^{(k)}_i(s,u)
  + \sum_{k,i} A_{i}^{(k)} Q^{(k)}_i(s,u)
    \bigg) \,,
\ee
where
\be
\begin{split}
\sum_{k,i}  r_i^{(k)} P^{(k)}_i(s,u)
 =~& r_0^{(0)}  
 + 
   r_1^{(2)} (s^2 + u^2)
   + r_2^{(2)} s u 
   + r_3^{(3)} s u (s + u)  
 + r_5^{(4)} s^2 u^2 + 
\\
&
    r_4^{(4)} (s^4 + u^4)
    +  
    r_6^{(5)} s u(s^3 + u^3)
    +r_7^{(5)} s^2 u^2 (s + u)   + 
 \ldots
 \end{split}
 \label{monoexp}
\ee
and  
{\footnotesize
\begin{eqnarray}
\nonumber
&&\!\hspace{-0.8cm}\sum_{k,i} A_{i}^{(k)} Q^{(k)}_i(s,u)
 \\ \nonumber
 =&&
 a_{1,0} (s+u)
 \bigg[
    1 - \zeta_2 s u
    - \zeta_4 s u
    (s^2 + \tfrac{1}{4} s u + u^2)
    -\zeta_6 s u ( s^4 - s^3 u - \tfrac{33}{16} s^2 u^2 - s u^3 + u^4)+\ldots
 \bigg]
 \\ \nonumber
 &&
 + a_{3,0} (s+u)
 \bigg[
    (s^2 + s u + u^2) 
    - \zeta_2 s u (s^2 + s u + u^2) 
    - 
  \zeta_4 s u 
  ( s^4 - s^3 u - \tfrac{9}{4} s^2 u^2 - s u^3 + u^4)+ 
  \ldots
 \bigg]
  \\ \nonumber
   &&
   + a_{4,1} s u (s+u)^2
 \bigg[
    1
    - \zeta_2 s u 
    - \zeta_4 s u 
  (s^2 + \tfrac{1}{4} s u + u^2)+\ldots
 \bigg]
   \\ \nonumber
   &&
   + a_{5,0} (s+u)
 \bigg[
    (s^2 + s u + u^2)^2
    - \zeta_2 s u (s^4 - s^3 u - 3 s^2 u^2 - s u^3 + u^4)
    +\ldots
 \bigg]
    \\ \nonumber
   &&
   + a_{6,1} s u (s + u)^2
 \bigg[
    (s^2 + s u + u^2)
    - \zeta_2 s u (s^2 + s u + u^2)
    +\ldots
 \bigg]
  \\ 
  && +\ldots
  \label{beast}
\end{eqnarray}
}%
The coefficients in the expression \reef{beast} can be shifted while preserving the Mandelstam  polynomial of lowest degree in each $Q^{(k)}_i(s,u)$. For example, taking $a_{3,0} \to \tilde{a}_{3,0} -\zeta_2 a_{1,0}$ changes the term $-\zeta_2 su$ in the first line of \reef{beast} to $-\zeta_2 (s+u)^2 = -\zeta_2 t^2$ while also modifying  higher powers in the Mandelstams multiplying $a_{1,0}$. Next, take $a_{5,0} \to \tilde{a}_{5,0} -\frac{3}{4} \zeta_4 a_{1,0}$ to make the $\zeta_4$-terms in the first line of  \reef{beast} only depend on $t$. Doing this repeatedly, we find that the series of Mandelstam terms that multiply $a_{1,0}$ only depends on $t$. Moreover, the terms are easily recognized as those in the series expansion of $\sin(\pi t)/\pi$. Thus, after these basic reparametrizations, we find evidence that the coefficient of  $a_{1,0}$ resums to $\sin(\pi t)/\pi$.  
Remarkably, a similar set of shifts also works for the higher-order coefficients $\tilde{a}_{3,0}$, $\tilde{a}_{4,1}$ etc, bringing each of them to a form that can be resummed to $\sin(\pi t)/\pi$ times a fully symmetric polynomial in $s,t,u$ of degree $k-1$. We have checked this explicitly to $k=20$ and find that
 \begin{eqnarray}
 \nonumber
&&\!\hspace{-0.8cm}\sum_{k,i} A_{i}^{(k)} Q^{(k)}_i(s,u)
\,=\, -\frac{1}{\pi}\sin(\pi t)
 \bigg[
    \tilde{a}_{1,0}
    + \tilde{a}_{3,0}\, \sigma_2
    + \tilde{a}_{4,1}\, \sigma_3
    + \tilde{a}_{5,0} \,
    \sigma_2^2
    + \tilde{a}_{6, 1} \, \sigma_2\sigma_3
    \\
    &&
    \hspace{4.9cm}
    + \tilde{a}_{7, 0}\,\sigma_2^3
    +  \tilde{a}_{7, 2}\, \sigma_3^2
  + \tilde{a}_{8, 1}\, \sigma_2^2\sigma_3+
    \ldots
 \bigg]\,,
  \label{beauty}
 \end{eqnarray}
where we have defined
\be
 \sigma_2 = \frac{1}{2}(s^2+t^2+u^2)
 ~~~\text{and}~~~
 \sigma_3 = - s t u
\ee
and the sum continues over all the independent Mandelstam polynomials $\sigma_2^{n}\sigma_3^{m}$ fully symmetric in $s,t,u$. The coefficients in \reef{beauty} are related to those in \reef{beast} via finite shifts:
\be
\label{coeffshifts}
 \begin{split}
 \tilde{a}_{1,0} =\,&a_{1,0}\,,\\
 \tilde{a}_{3,0} =\,& a_{3,0} +\zeta_2\, a_{1,0}\,,\\
 \tilde{a}_{4,1} =\,&a_{4,1}\,,\\
 \tilde{a}_{5,0} =\,& a_{5,0} 
 +\zeta_2\, a_{3,0}
  +\frac{7}{4} \zeta_4\, a_{1,0}\,,
  \\
 \tilde{a}_{6,1} =\,& a_{6,1} 
 +\zeta_2\, a_{4,1}\,,
 \\
 \tilde{a}_{7,0} =\,& a_{7,0} 
 +\zeta_2\, a_{5,0}
  +\frac{7}{4} \zeta_4\, a_{3, 0} 
  + \frac{31}{16} \zeta_6\, a_{1, 0}
  \,,\\
   \tilde{a}_{7,2} =\,& a_{7,2} 
   - 9a_{7,0}
 +3 \zeta_2\, a_{5,0}
  +\frac{9}{4} \zeta_4\, a_{3, 0} 
  + \frac{9}{4} \zeta_6\, a_{1, 0}\,,
  \\
   \tilde{a}_{8,1} =\,& a_{8,1} 
 +\zeta_2\, a_{6,1} 
 + \frac{7}{4} \zeta_4\, a_{4, 1} \,,~~~~
 \text{etc}
 \end{split}
\ee
The point of this different parameterization of the low-energy amplitude is that for any choice of monovariables $r_i^{(k)}$ in the allowed region, the large-$k_\text{max}$ limit of the S-matrix bootstrap will fix the coefficients $\tilde{a}_{k,q}$ in the partially resummed symmetric Mandelstam polynomial expression \reef{beauty}, as illustrated Figure \reef{fig:cartoon} and tested 
in examples in Section \ref{sec:emcodconj2}.  What this tell us about the UV theory remains a question for the future.

However, we can make some sense of the parameterization in \reef{newexpA4again} and \reef{beauty}. Recall that the motivation for the  monovariables came from the string monodromy relations \reef{monodromrel}. The Wilson coefficients that appear in $\sum_{k,i} A_{i}^{(k)} Q^{(k)}_i(s,u)$ are  those that were left unfixed by the monodromy relations. With that in mind, it is straighforward to see that any function $f$ of the form 
\be
 f(s,u) = \sin(\pi t) \,g(s,t,u)\,,~~~~
 s+t+u = 0
\ee
where $g$ is fully symmetric in $s,t,u$, solves the string monodromy relations \reef{monodromrelReIm}. Expanding $g$ in the most general polynomial form then gives precisely the partially resummed form of $\sum_{k,i} A_{i}^{(k)} Q^{(k)}_i(s,u)$ in  \reef{beauty}.

Including the general monovariables has nothing to do with string monodromy relations: their definition was inspired by the monodromy relations, but rather than having fixed values for the string (as given in Table \ref{monoconk4}) they can be free chosen in the allowed region and for general choices may not be any linear amplitudes relations associated with the change of variables.%
\footnote{The Veneziano amplitude can also be written as
\begin{align}
A^{\text{str}}[zz\bar{z}\bar{z}] = (\a's)^2\frac{\sin(\pi\a't)}{\pi}\G(-\a's)\G(-\a't)\G(-\a'u).
\end{align}
In this form, its low-energy expansion takes the form of \reef{beauty} with some particular set of $\tilde{a}_{k,q}'$ coefficients. That $\tilde{a}_{k,q}'$ basis  includes dependencies on $a_{k,q}$ Wilson coefficients that are fixed by monodromy such that the monovariables, some $\tilde{r}_{i}'^{(k)}$, which would appear in \reef{monoexp}, are all identically zero for the string. This is not the basis that we have chosen because we want to explicitly separate those values we fix with the monodromy constraints from those we do not.
}

\section{Discussion}
\label{s:discuss}
We have studied universal bounds in a simple theory: planar $\mathcal{N}=4$ SYM with higher-derivative corrections. Supersymmetry allows us to derive dispersive representations for all nonzero Wilson coefficients and the resulting bounds are studied using numerical implementation in SDPB and CPLEX. 

A key finding in this paper is the evidence that the EFT-hedron flattens out in the $k_\text{max} \to \infty$ limit. This leads us to conjecture that imposing positivity and fixing two-thirds of the Wilson coefficients (asymptotically) fixes the remaining third of Wilson coefficients. We numerically cross-checked the conjecture for a collection of randomly generated theories in Section \ref{sec:emcodconj2}. One moral of this story is that bounds on large numbers of Wilson coefficients are significantly stronger than one might naively expect. It would be interesting to understand what this phenomenon tells us about the UV theory. 
The novel, partially resummed expansion of the low-energy amplitude may be a step in that direction. 

Our analysis shares many features  with the pion-bootstrap papers \cite{Albert:2022oes,Fernandez:2022kzi}.  One difference is that supersymmetry effectively imposes additional constraints and allows us to bound all nonzero Wilson coefficients without having to assume a stronger Froissart bound. 
As an example of similar results, 
Figure 1 of \cite{Albert:2022oes} shares qualitative features with our $(\bar{a}_{2,0},\bar{a}_{2,1})$ plot in Figure \ref{fig:a20a21proj}.

In the Introduction, we mentioned that the input of string monodromy can be reframed as the assumption that the $\mathcal{N}=4$ SYM EFT amplitudes are obtained from a double copy \reef{N4dcEFT}. 
Let us elaborate on how this can be seen as more of a ``pure field theory'' input. 

Consider the double-copy
\be\label{dcEFT}
\begin{split}
 \text{(YM EFT)} 
 & =
 \text{(BAS EFT)}
\otimes_\text{FT}
 \text{(pure YM)} \,,\\
 A^\text{YM EFT}_n[\alpha] 
 & = \sum_{\beta,\gamma} 
  m_n[\alpha|\beta] \,S_n[a|\gamma] \,A_n^\text{YM}[\gamma] \, ,
 \end{split}
\ee
 where BAS EFT stands for a general Bi-Adjoint Scalar (BAS) model with local higher-derivative interactions and the subscript ``FT'' indicates that double-copy is done in the field theory limit; i.e.~the double-copy kernel $S_n[\beta|\gamma]$  adds no higher-derivative terms. 
The BAS EFT tree amplitudes are doubly-color ordered, $m_n[\alpha|\beta]$. In the double-copy \reef{dcEFT}, a choice of $(n-3)!$ out of $(n-1)!$ inequivalent color-orderings for $\beta$ and $\gamma$ must be summed over. In order for the result, $A^\text{YM EFT}_n$, to be independent of these choices (and thereby validate the double-copy as a map between field theories), it was argued in  \cite{Chi:2021mio}
that the amplitudes $m_n[\alpha|\beta]$ must obey the same linear relations on the second color-structure as the pure YM amplitudes, namely the Kleiss-Kuijf (KK) and Bern-Carrasco-Johansson (BCJ) field theory relations. 
It was shown in \cite{Chen:2022shl} that imposing, as required above, the KK and BCJ relations on the second color-ordering of $m_n[\alpha|\beta]$ implies (as checked to 36th order the derivative expansion at 4-point)  that the $m_n[\alpha|\beta]$ amplitude obeys a second set of linear relations on the first color-ordering that are highly constrained by locality. These new relations can be identified as the low-energy expansion of the monodromy relations! Since the YM EFT amplitude $A^\text{YM EFT}_n[\alpha]$ inherits its color-structure from $m_n[\alpha|\beta]$, we conclude that any amplitude constructed this way necessarily obeys the low-energy expansion of the monodromy relations.

Adding $\mathcal{N}=4$ supersymmetry to the double-copy \reef{dcEFT} gives \reef{N4dcEFT} and, putting all the above information together, we arrive at the claim stated in the Introduction: among the 4-point amplitudes that arise from \reef{N4dcEFT}, the unique one that is compatible with  unitarity, locality, and the Froissart bound is the Veneziano open string tree amplitude.

There are a number of possible future directions. Of particular interest would be to investigate how one might isolate string theory in the $\mathcal{N} = 4$ supersymmetric EFT-hedron in a more generic way. Other than imposing monodromy or that the low-energy amplitude must satisfy some double-copy constraints, there could be purely physical assumptions one can add that either uniquely pick out the tree-level open string amplitude or clearly place the string at a corner of the allowed region. Finding such physical assumptions would provide insight into what distinguishes string theory, at least from the low-energy perspective.  
 
We found evidence that fixing about two-thirds of the Wilson coefficients as monovariables is sufficient to show that the EFT-hedron flattens in the $k_{\max}\to \infty$ limit. It is possible that fixing just some subset of the monovariables is enough to fix all other coefficients. Finding the lower bound on the number of monovariables would reveal the most efficient parameterization of the flattened EFT-hedron.

It is unclear how generic the flattening phenomenon is. For example, it would be interesting to examine whether flattening also occurs in the pion bootstrap \cite{Guerrieri_2021,Zahed:2021fkp,Albert:2022oes,Fernandez:2022kzi,Albert:2023jtd,he2023bootstrapping} or for abelian scalar models as in \cite{Caron-Huot:2020cmc}. A theory with reduced or no supersymmetry has more independent Wilson coefficients and the positivity bounds may be significantly more complex as well. Therefore, one could also examine whether flattening occurs in $\mathcal{N}=8$ supergravity. In our analysis, the monodromy relations were helpful for identifying which directions the flattening happens along.  
The challenge of studying flattening in other cases, especially those without color-structure, is that there is no obvious candidate for a replacement of monodromy relations.


\section*{Acknowledgements}
We would like to thank Jan Albert, Enrico Hermann, Loki Lin, Andrew Neitzke, Leonardo Rastelli, David Poland, and Nick Geiser
for useful comments and discussions. HE and JB are supported in part by DE-SC0007859. AH was supported in part by a Rackham Predoctoral Fellowship from the University of Michigan and in part by the Simons Foundation.

\appendix

\section{Convergence of Numerical Results}\label{app:linprogacc}
The semi-definite and linear programming algorithms described in Sections \ref{sec:sdpb} and \ref{sec:CPLEX} approximate the EFT-hedron from the inside due to the truncation in $\ell$ and, in the case of linear programming, $x_{\max}$. 
In Section \ref{s:CPLEXvsSDPB}, we discussed how CPLEX approaches the SDPB results as we increase $x_{\max}$ for  given $k_\text{max}$ and $\ell_{\max}$. 
In this Appendix, we illustrate the dependence of the SDPB results on $\ell_{\max}$ and describe how we find accurate bounds on Wilson coefficients.

\subsection{Convergence without Monodromy}

Because of the spin cut-off $\ell_\text{max}$, the bounds on ratios of Wilson coefficients are approximated from the inside of the allowed region for given fixed $k_\text{max}$. With higher $k_\text{max}$, i.e.~more null constraints, it is necessary to increase $\ell_{\max}$ to get accurate results. 
To find the necessary $\ell_{\max}$, we compute the bound for a given Wilson coefficient by increasing $\ell_{\max}$ until we obtain convergence as a function of $\ell_{\max}$. The computation time typically increases linearly with $\ell_\text{max}$.  Practically, for our plots we use the value of $\ell_\text{max}$ that matches the asymptotic bound with the precision needed.

\begin{figure}
\centering
\includegraphics[width=0.6\textwidth]{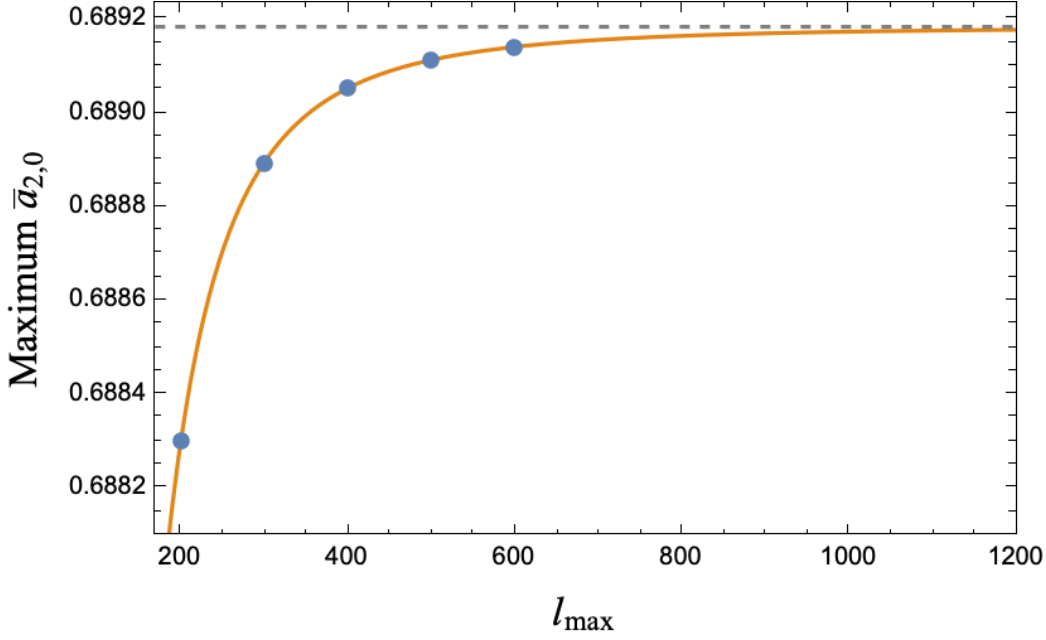}
\caption{\label{fig:asyanala20a21proj} Values of the maximal $a_{2,0}/a_{0,0}$ at $k_{\max}=20$ when $a_{2,1}/a_{0,0}$ is fixed to its string value  from points with $\ell_{\max}$ between 200 and 600. The orange curve is a fit of these points as a power law and the asymptotic value is given as a gray, dashed line.}
\end{figure}

 For example, in Fig. \ref{fig:asyanala20a21proj}, we stop computing the maximal values for $a_{2,0}$ at $\ell_{\max} = 600$. We fit the points up to $\ell_{\max}$ to a power law function
\begin{align}
a_{2,0}^{\max}(\ell_{\max}) = \frac{A}{\ell_{\max}^\g}+b
\end{align}
to find the asymptotic value, $b$. When we plot the point with $a_{2,0} = b$ at the string value on the right hand side of Fig. \ref{fig:a20a21proj}, there is no visual difference to the value computed at $\ell_{\max} = 600$. In this sense, the $\ell_{\max} = 600$ value is precise enough for our plots. This is the general technique we use to determine the minimal $\ell_{\max}$ to use when computing bounds.

\subsection{Convergence with Monodromy}
\label{app:smonod}

\begin{figure}
\centering
\includegraphics[width=0.6\textwidth]{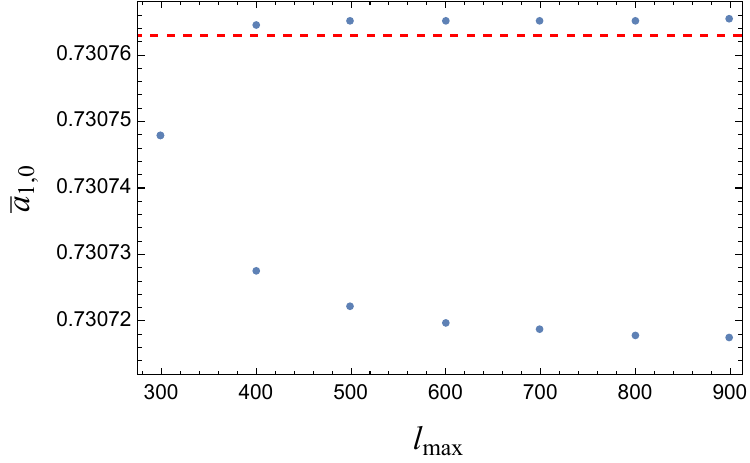}~~
\caption{\label{fig:a120lmax8} Plots of upper and lower bounds on $a_{1,0}$ at $k_{\max}=8$ with monodromy constraints. The red dashed line is the string value for $\bar{a}_{1,0}$}
\end{figure}

\begin{figure}
\centering
\includegraphics[width=0.4\textwidth]{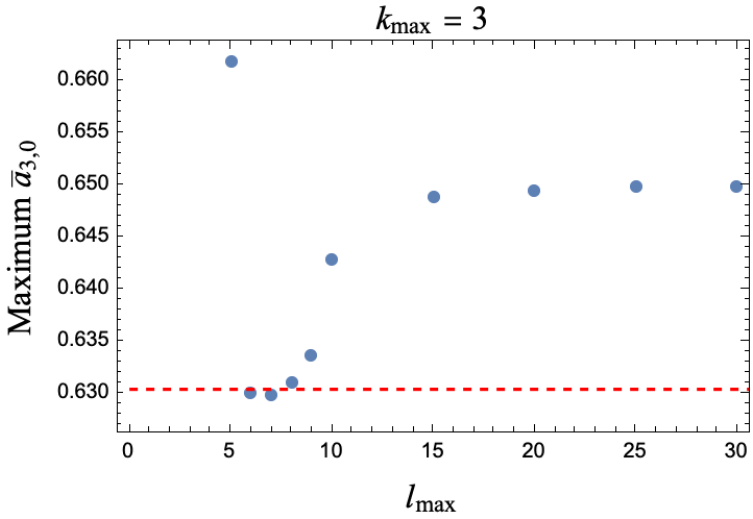}~~\includegraphics[width=0.4\textwidth]{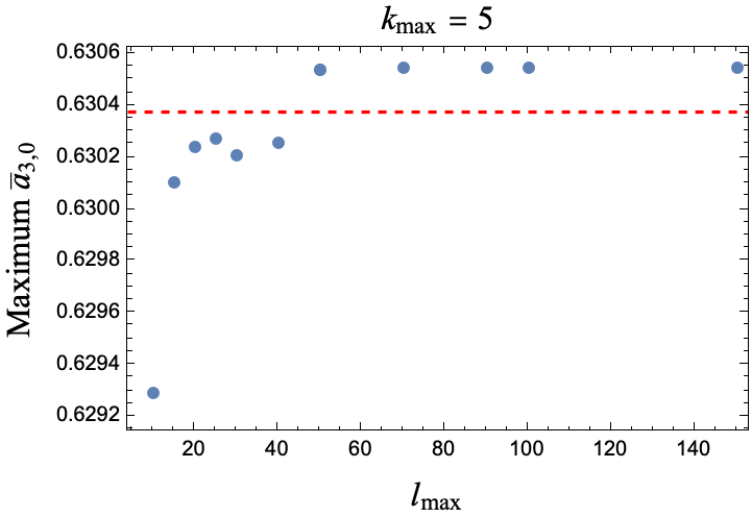} \\
\includegraphics[width=0.4\textwidth]{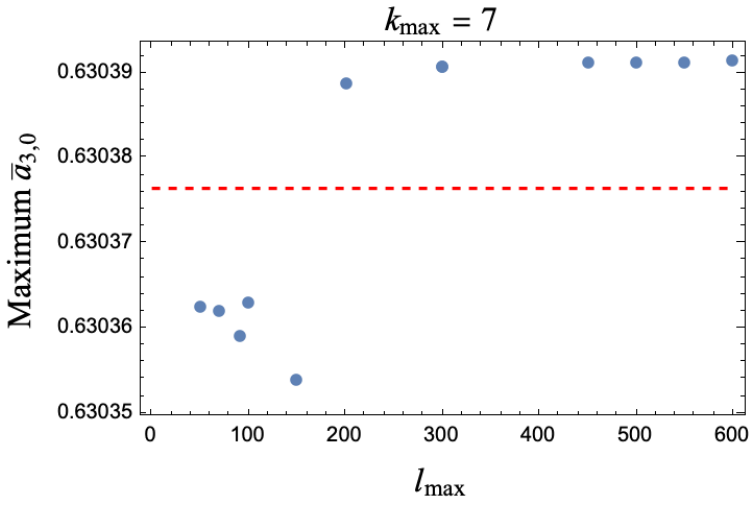}~~\includegraphics[width=0.4\textwidth]{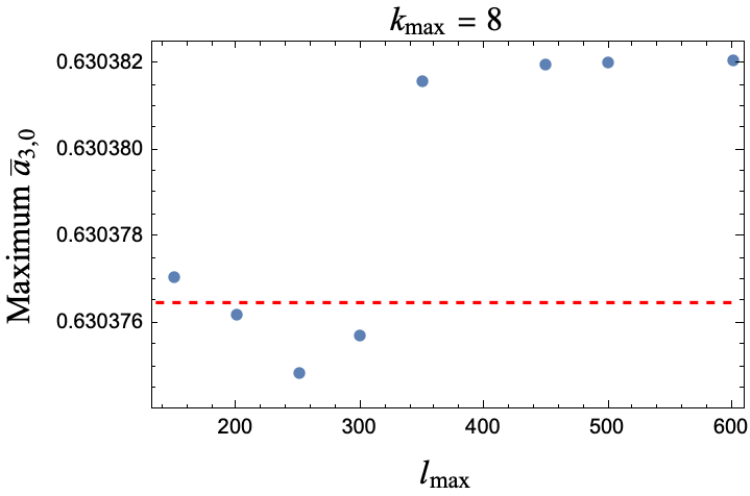}
\caption{\label{fig:a30lmax8} Plots of the upper bound on $a_{3,0}$ as a function of $\ell_{\max}$ at $k_{\max}=3,5,7,8$ with monodromy constraints. The red line indicates the string  value which necessarily lies below the value the upper bound computed by SDPB.}
\end{figure}

Adding in monodromy constraints increases the needed $\ell_{\max}$.  Figure  \ref{fig:a120lmax8}, shows a plot of the maximum and minimum of $a_{1,0}$ given by SDPB as a function of $\ell_{\max}$ at $k_{\max}=8$ with monodromy constraints imposed. It is clear that the lower and upper bounds converge starting around $\ell_{\max}=800$ and $\ell_{\max}=500$ respectively. Other plots for upper bounds on $a_{3,0}$ for $k_{\max}=3,5,7,8$ are given in Figure  \ref{fig:a30lmax8}. These plots show how too-low $\ell_\text{max}$ yields nonsensical non-monotonically increasing results. Therefore, for each $k_\text{max}$, one needs to take $\ell_\text{max}$ large enough to get extremization values that converge properly. 
We cross-checked results between CPLEX and SDPB at higher $\ell_{\max}$ and found agreement. As discussed in Sections \reef{sec:CPLEX} and \reef{s:CPLEXvsSDPB}, for CPLEX, one similarly has to benchmark the fineness of discretization, $x_\text{max}$.

\bibliographystyle{JHEP}
\bibliography{Draft.bib}

\providecommand{\href}[2]{#2}\begingroup\raggedright\begin{thebibliography}{10}

\bibitem{Huang:2020nqy}
Y.-t. Huang, J.-Y. Liu, L.~Rodina and Y.~Wang, \emph{{Carving out the Space of
  Open-String S-matrix}},
  \href{https://doi.org/10.1007/JHEP04(2021)195}{\emph{JHEP} {\bfseries 04}
  (2021) 195} [\href{https://arxiv.org/abs/2008.02293}{{\ttfamily
  2008.02293}}].

\bibitem{Arkani-Hamed:2020blm}
N.~Arkani-Hamed, T.-C. Huang and Y.-T. Huang, \emph{{The EFT-Hedron}},
  \href{https://doi.org/10.1007/JHEP05(2021)259}{\emph{JHEP} {\bfseries 05}
  (2021) 259} [\href{https://arxiv.org/abs/2012.15849}{{\ttfamily
  2012.15849}}].

\bibitem{Caron-Huot:2020cmc}
S.~Caron-Huot and V.~Van~Duong, \emph{{Extremal Effective Field Theories}},
  \href{https://arxiv.org/abs/2011.02957}{{\ttfamily 2011.02957}}.

\bibitem{Chiang:2021ziz}
L.-Y. Chiang, Y.-t. Huang, W.~Li, L.~Rodina and H.-C. Weng, \emph{{Into the
  EFThedron and UV constraints from IR consistency}},
  \href{https://doi.org/10.1007/JHEP03(2022)063}{\emph{JHEP} {\bfseries 03}
  (2022) 063} [\href{https://arxiv.org/abs/2105.02862}{{\ttfamily
  2105.02862}}].

\bibitem{Guerrieri_2021}
A.~L. Guerrieri, J.~Penedones and P.~Vieira, \emph{S-matrix bootstrap for
  effective field theories: massless pions},
  \href{https://doi.org/10.1007/jhep06(2021)088}{\emph{Journal of High Energy
  Physics} {\bfseries 2021} (2021) }.

\bibitem{Zahed:2021fkp}
A.~Zahed, \emph{{Positivity and geometric function theory constraints on pion
  scattering}}, \href{https://doi.org/10.1007/JHEP12(2021)036}{\emph{JHEP}
  {\bfseries 12} (2021) 036}
  [\href{https://arxiv.org/abs/2108.10355}{{\ttfamily 2108.10355}}].

\bibitem{Albert:2022oes}
J.~Albert and L.~Rastelli, \emph{{Bootstrapping pions at large N}},
  \href{https://doi.org/10.1007/JHEP08(2022)151}{\emph{JHEP} {\bfseries 08}
  (2022) 151} [\href{https://arxiv.org/abs/2203.11950}{{\ttfamily
  2203.11950}}].

\bibitem{Fernandez:2022kzi}
C.~Fernandez, A.~Pomarol, F.~Riva and F.~Sciotti, \emph{{Cornering Large-$N_c$
  QCD with Positivity Bounds}},
  \href{https://arxiv.org/abs/2211.12488}{{\ttfamily 2211.12488}}.

\bibitem{Albert:2023jtd}
J.~Albert and L.~Rastelli, \emph{{Bootstrapping Pions at Large $N$. Part II:
  Background Gauge Fields and the Chiral Anomaly}},
  \href{https://arxiv.org/abs/2307.01246}{{\ttfamily 2307.01246}}.

\bibitem{he2023bootstrapping}
Y.~He and M.~Kruczenski, \emph{Bootstrapping gauge theories},
  \href{https://arxiv.org/abs/2309.12402}{{\ttfamily 2309.12402}}.

\bibitem{Alberte:2020bdz}
L.~Alberte, C.~de~Rham, S.~Jaitly and A.~J. Tolley, \emph{{QED positivity
  bounds}}, \href{https://doi.org/10.1103/PhysRevD.103.125020}{\emph{Phys. Rev.
  D} {\bfseries 103} (2021) 125020}
  [\href{https://arxiv.org/abs/2012.05798}{{\ttfamily 2012.05798}}].

\bibitem{Henriksson:2021ymi}
J.~Henriksson, B.~McPeak, F.~Russo and A.~Vichi, \emph{{Rigorous bounds on
  light-by-light scattering}},
  \href{https://doi.org/10.1007/JHEP06(2022)158}{\emph{JHEP} {\bfseries 06}
  (2022) 158} [\href{https://arxiv.org/abs/2107.13009}{{\ttfamily
  2107.13009}}].

\bibitem{Haring:2022sdp}
K.~H\"aring, A.~Hebbar, D.~Karateev, M.~Meineri and J.~a. Penedones,
  \emph{{Bounds on photon scattering}},
  \href{https://arxiv.org/abs/2211.05795}{{\ttfamily 2211.05795}}.

\bibitem{deRham:2022sdl}
C.~de~Rham, L.~Engelbrecht, L.~Heisenberg and A.~L\"uscher, \emph{{Positivity
  bounds in vector theories}},
  \href{https://doi.org/10.1007/JHEP12(2022)086}{\emph{JHEP} {\bfseries 12}
  (2022) 086} [\href{https://arxiv.org/abs/2208.12631}{{\ttfamily
  2208.12631}}].

\bibitem{Chowdhury:2021ynh}
S.~D. Chowdhury, K.~Ghosh, P.~Haldar, P.~Raman and A.~Sinha, \emph{{Crossing
  Symmetric Spinning S-matrix Bootstrap: EFT bounds}},
  \href{https://doi.org/10.21468/SciPostPhys.13.3.051}{\emph{SciPost Phys.}
  {\bfseries 13} (2022) 051}
  [\href{https://arxiv.org/abs/2112.11755}{{\ttfamily 2112.11755}}].

\bibitem{Caron-Huot:2022ugt}
S.~Caron-Huot, Y.-Z. Li, J.~Parra-Martinez and D.~Simmons-Duffin,
  \emph{{Causality constraints on corrections to Einstein gravity}},
  \href{https://arxiv.org/abs/2201.06602}{{\ttfamily 2201.06602}}.

\bibitem{Simmons-Duffin:2015qma}
D.~Simmons-Duffin, \emph{{A Semidefinite Program Solver for the Conformal
  Bootstrap}}, \href{https://doi.org/10.1007/JHEP06(2015)174}{\emph{JHEP}
  {\bfseries 06} (2015) 174}
  [\href{https://arxiv.org/abs/1502.02033}{{\ttfamily 1502.02033}}].

\bibitem{Landry:2019qug}
W.~Landry and D.~Simmons-Duffin, \emph{{Scaling the semidefinite program solver
  SDPB}},  \href{https://arxiv.org/abs/1909.09745}{{\ttfamily 1909.09745}}.

\bibitem{cplex2009v12}
I.~I. Cplex, \emph{V12. 1: User’s manual for cplex}, {\emph{International
  Business Machines Corporation} {\bfseries 46} (2009) 157}.

\bibitem{Plahte:1970wy}
E.~Plahte, \emph{{Symmetry properties of dual tree-graph n-point amplitudes}},
  \href{https://doi.org/10.1007/BF02824716}{\emph{Nuovo Cim. A} {\bfseries 66}
  (1970) 713}.

\bibitem{Stieberger:2009hq}
S.~Stieberger, \emph{{Open \textbackslash{}\& Closed vs. Pure Open String Disk
  Amplitudes}},  \href{https://arxiv.org/abs/0907.2211}{{\ttfamily 0907.2211}}.

\bibitem{Bjerrum-Bohr:2009ulz}
N.~E.~J. Bjerrum-Bohr, P.~H. Damgaard and P.~Vanhove, \emph{{Minimal Basis for
  Gauge Theory Amplitudes}},
  \href{https://doi.org/10.1103/PhysRevLett.103.161602}{\emph{Phys. Rev. Lett.}
  {\bfseries 103} (2009) 161602}
  [\href{https://arxiv.org/abs/0907.1425}{{\ttfamily 0907.1425}}].

\bibitem{Bjerrum-Bohr:2010mia}
N.~E.~J. Bjerrum-Bohr, P.~H. Damgaard, T.~Sondergaard and P.~Vanhove,
  \emph{{Monodromy and Jacobi-like Relations for Color-Ordered Amplitudes}},
  \href{https://doi.org/10.1007/JHEP06(2010)003}{\emph{JHEP} {\bfseries 06}
  (2010) 003} [\href{https://arxiv.org/abs/1003.2403}{{\ttfamily 1003.2403}}].

\bibitem{Bjerrum-Bohr:2010pnr}
N.~E.~J. Bjerrum-Bohr, P.~H. Damgaard, T.~Sondergaard and P.~Vanhove,
  \emph{{The Momentum Kernel of Gauge and Gravity Theories}},
  \href{https://doi.org/10.1007/JHEP01(2011)001}{\emph{JHEP} {\bfseries 01}
  (2011) 001} [\href{https://arxiv.org/abs/1010.3933}{{\ttfamily 1010.3933}}].

\bibitem{Chen:2022shl}
A.~S.-K. Chen, H.~Elvang and A.~Herderschee, \emph{{Emergence of String
  Monodromy in Effective Field Theory}},
  \href{https://arxiv.org/abs/2212.13998}{{\ttfamily 2212.13998}}.

\bibitem{yutinetal}
L.-Y. Chiang, Y.~tin Huang and H.-C. Weng, ``Bootstrapping string theory
  eft.''.

\bibitem{Elvang:2013cua}
H.~Elvang and Y.-t. Huang, \emph{{Scattering Amplitudes}},
  \href{https://arxiv.org/abs/1308.1697}{{\ttfamily 1308.1697}}.

\bibitem{Elvang:2015rqa}
H.~Elvang and Y.-t. Huang, \emph{{Scattering Amplitudes in Gauge Theory and
  Gravity}}. Cambridge University Press, 4, 2015.

\bibitem{Maity:2021obe}
P.~Maity, \emph{{Positivity of the Veneziano amplitude in D = 4}},
  \href{https://doi.org/10.1007/JHEP04(2022)064}{\emph{JHEP} {\bfseries 04}
  (2022) 064} [\href{https://arxiv.org/abs/2110.01578}{{\ttfamily
  2110.01578}}].

\bibitem{Arkani-Hamed:2022gsa}
N.~Arkani-Hamed, L.~Eberhardt, Y.-t. Huang and S.~Mizera, \emph{{On unitarity
  of tree-level string amplitudes}},
  \href{https://doi.org/10.1007/JHEP02(2022)197}{\emph{JHEP} {\bfseries 02}
  (2022) 197} [\href{https://arxiv.org/abs/2201.11575}{{\ttfamily
  2201.11575}}].

\bibitem{Craig:2011ws}
N.~Craig, H.~Elvang, M.~Kiermaier and T.~Slatyer, \emph{{Massive amplitudes on
  the Coulomb branch of N=4 SYM}},
  \href{https://doi.org/10.1007/JHEP12(2011)097}{\emph{JHEP} {\bfseries 12}
  (2011) 097} [\href{https://arxiv.org/abs/1104.2050}{{\ttfamily 1104.2050}}].

\bibitem{Kiermaier:2011cr}
M.~Kiermaier, \emph{{The Coulomb-branch S-matrix from massless amplitudes}},
  \href{https://arxiv.org/abs/1105.5385}{{\ttfamily 1105.5385}}.

\bibitem{Herderschee:2019dmc}
A.~Herderschee, S.~Koren and T.~Trott, \emph{{Constructing $ \mathcal{N} $ = 4
  Coulomb branch superamplitudes}},
  \href{https://doi.org/10.1007/JHEP08(2019)107}{\emph{JHEP} {\bfseries 08}
  (2019) 107} [\href{https://arxiv.org/abs/1902.07205}{{\ttfamily
  1902.07205}}].

\bibitem{Boels:2010mj}
R.~H. Boels, \emph{{No triangles on the moduli space of maximally
  supersymmetric gauge theory}},
  \href{https://doi.org/10.1007/JHEP05(2010)046}{\emph{JHEP} {\bfseries 05}
  (2010) 046} [\href{https://arxiv.org/abs/1003.2989}{{\ttfamily 1003.2989}}].

\bibitem{Abhishek:2023lva}
M.~Abhishek, S.~Hegde, D.~P. Jatkar, A.~P. Saha and A.~Suthar, \emph{{Loop
  Amplitudes in the Coulomb Branch of $\mathcal{N}=4$ Super-Yang-Mills
  Theory}},  \href{https://arxiv.org/abs/2308.05705}{{\ttfamily 2308.05705}}.

\bibitem{Davydychev:1993ut}
A.~I. Davydychev, \emph{{Standard and hypergeometric representations for loop
  diagrams and the photon-photon scattering}},  in \emph{{7th International
  Seminar on High-energy Physics}}, 5, 1993,
  \href{https://arxiv.org/abs/hep-ph/9307323}{{\ttfamily hep-ph/9307323}}.

\bibitem{Camanho:2014apa}
X.~O. Camanho, J.~D. Edelstein, J.~Maldacena and A.~Zhiboedov, \emph{{Causality
  Constraints on Corrections to the Graviton Three-Point Coupling}},
  \href{https://doi.org/10.1007/JHEP02(2016)020}{\emph{JHEP} {\bfseries 02}
  (2016) 020} [\href{https://arxiv.org/abs/1407.5597}{{\ttfamily 1407.5597}}].

\bibitem{Heisenberg:1949kqa}
W.~Heisenberg, \emph{{Production of Meson Showers}},
  \href{https://doi.org/10.1038/164065c0}{\emph{Nature} {\bfseries 164} (1949)
  65}.

\bibitem{Froissart:1961ux}
M.~Froissart, \emph{{Asymptotic behavior and subtractions in the Mandelstam
  representation}}, \href{https://doi.org/10.1103/PhysRev.123.1053}{\emph{Phys.
  Rev.} {\bfseries 123} (1961) 1053}.

\bibitem{Chi:2021mio}
H.-H. Chi, H.~Elvang, A.~Herderschee, C.~R.~T. Jones and S.~Paranjape,
  \emph{{Generalizations of the double-copy: the KLT bootstrap}},
  \href{https://doi.org/10.1007/JHEP03(2022)077}{\emph{JHEP} {\bfseries 03}
  (2022) 077} [\href{https://arxiv.org/abs/2106.12600}{{\ttfamily
  2106.12600}}].

\end{thebibliography}\endgroup

\end{document}